\documentclass[journal, romanappendices]{IEEEtran}
\ifCLASSINFOpdf
\else
\fi
\usepackage[utf8]{inputenc} 
\usepackage[T1]{fontenc}
\usepackage{url}
\usepackage{ifthen}
\usepackage{cite}
\usepackage{graphicx}
\usepackage{hyperref}
\usepackage{amsmath}
\usepackage{amssymb}
\usepackage{bm}
\usepackage{bbm}
\usepackage{algorithmicx}
\usepackage{algorithm}
\usepackage{algpseudocode}
\usepackage{array}
\usepackage{booktabs}
\usepackage{multirow}
\usepackage{makecell}
\usepackage{mathtools}
\usepackage{mathrsfs}
\usepackage{amsthm}
\usepackage{xcolor}

\newtheorem{definition}{Definition}
\newtheorem{theorem}{Theorem}
\newtheorem{remark}{Remark}
\newtheorem{lemma}{Lemma}

\newtheorem{corollary}{Corollary}

\newtheorem{fact}{Fact}

\newcommand{\X}{\mathcal{X}}
\newcommand{\Y}{\mathcal{Y}}
\newcommand{\E}{\mathbb{E}}
\newcommand{\R}{\mathbb{R}}
\newcommand{\F}{\mathcal{F}}
\newcommand{\Prob}{\mathscr{P}}
\newcommand{\indicator}{\mathbf{1}}
\newcommand{\setS}{\mathcal{S}}

\newcommand{\Brace}[1]{\left\{#1\right\}}
\newcommand{\Paren}[1]{\left(#1\right)}
\newcommand{\Br}[1]{\left[#1\right]}
\newcommand*\diff{\mathop{}\!\mathrm{d}}

\DeclareMathOperator*{\argmax}{arg\,max}
\DeclareMathOperator*{\argmin}{arg\,min}

\begin{document}
%
\title{Sequential Transmission Over Binary Asymmetric Channels With Feedback}
%
%
%

\author{Hengjie~Yang,~\IEEEmembership{Student Member,~IEEE},
        Minghao~Pan,
        Amaael~Antonini,
        and~Richard~D.~Wesel,~\IEEEmembership{Fellow,~IEEE}
\thanks{
This paper was presented in part at 2020 IEEE International Symposium on Information Theory (ISIT) \cite{Yang2020}. 

This research is supported by National Science Foundation (NSF) grant CCF-1955660. Any opinions, findings, and conclusions or recommendations expressed in this material  are those of the author(s) and do not necessarily reflect views of the NSF.

H.~Yang is with the Department of Electrical and Computer Engineering, University of California, Los Angeles, Los Angeles, CA, 90095 USA (e-mail: hengjie.yang@ucla.edu).

M.~Pan is with the Department of Mathematics, University of California, Los Angeles, Los Angeles, CA, 90095 USA (e-mail: minghaopan@ucla.edu).

A.~Antonini is with the Department of Electrical and Computer Engineering, University of California, Los Angeles, Los Angeles, CA, 90095 USA (e-mail: amaael@ucla.edu).

R.~D.~Wesel is with the Department of Electrical and Computer Engineering, University of California, Los Angeles, Los Angeles, CA, 90095 USA (e-mail: wesel@ucla.edu).
}}

\maketitle

\begin{abstract}
In this paper, we consider the problem of variable-length coding over the class of memoryless binary asymmetric channels (BACs) with noiseless feedback, including the binary symmetric channel (BSC) as a special case. In 2012, Naghshvar \emph{et al.} introduced an encoding scheme, which we refer to as the small-enough-difference (SED) encoder, which asymptotically achieves both capacity and Burnashev's optimal error exponent for symmetric binary-input channels. Building on the work of Naghshvar \emph{et al.}, this paper extends the SED encoding scheme to the class of BACs and develops a non-asymptotic upper bound on the average blocklength that is shown to achieve both capacity and the optimal error exponent. For the specific case of the BSC, we develop an additional non-asymptotic bound using a two-phase analysis that leverages both a submartingale synthesis and a Markov chain time of first passage analysis. For the BSC with capacity $1/2$,  both new achievability bounds exceed the achievability bound of Polyanskiy \emph{et al.} for a system limited to stop-feedback codes.
\end{abstract}

\begin{IEEEkeywords}
Binary asymmetric channels, variable-length coding, Burnashev's optimal error exponent, submartingales.
\end{IEEEkeywords}

%
\IEEEpeerreviewmaketitle

\section{Introduction}
%
%
%
%
\IEEEPARstart{F}{eedback} does not increase the capacity of memoryless channels \cite{Shannon1956}, but it can significantly reduce the complexity of communication and the probability of error, provided that variable-length feedback (VLF) codes are allowed. In the context of a discrete memoryless channel (DMC) with noiseless feedback, Burnashev \cite{Burnashev1976} proposed a pioneering two-phase transmission scheme that obtains the \emph{exact} optimal error exponent for all rates below capacity. The first phase is called the \emph{communication phase}, during which the transmitter seeks to increase the receiver's posterior probability for the transmitted message.  The system transitions from the communication phase to the \emph{confirmation phase} when the largest posterior at the receiver exceeds a certain threshold $\zeta$.  
During the confirmation phase, two most distinguishable input symbols are used: one for the message with the largest posterior, and the other for the rest of messages. The confirmation phase continues until either the transmission terminates or the system returns to the communication phase. This two-phase encoder allows Burnashev to obtain an upper bound on the average blocklength that coincides asymptotically with the converse bound, thus producing the optimal error exponent. However, Burnashev did not provide an explicit non-asymptotic bound on the average blocklength for the DMC.

For the binary symmetric channel (BSC) with noiseless feedback, Horstein \cite{Horstein1963} developed a simple, one-phase scheme that maps each message to a subinterval in $[0, 1]$. The transmitter sends a $0$ if the subinterval of the true message lies entirely beneath the median and a $1$ if it lies entirely above the median. If the subinterval includes the median point, which will eventually happen as the subinterval of the highest posterior grows, then randomized encoding is employed. Horstein did not provide a rigorous proof to show that his scheme achieves capacity. In \cite{Burnashev1974}, Burnashev and Zigangirov showed that Horstein's scheme achieves the capacity of the BSC in the fixed blocklength setting. In \cite{Shayevitz2011}, Shayevitz and Feder generalized Horstein's scheme to the concept of posterior matching, thus validating the capacity-achieving property of Horstein's scheme in the variable-length setting. Since Horstein's work, several authors, e.g., \cite{Schalkwijk1971,Schalkwijk1973,Tchamkerten2002,Tchamkerten2006}, have constructed coding schemes for the BSC with noiseless feedback under various assumptions in order to attain capacity or Burnashev's optimal error exponent. 

Error exponent analysis of variable-length coding typically focuses on asymptotically long average blocklength at a fixed rate. In contrast, Polyanskiy \emph{et al.} \cite{Polyanskiy2011} showed that in the non-asymptotic regime, variable-length coding with noiseless feedback can provide a significant advantage in achievable rate over fixed-length codes. Polyanskiy \emph{et al.} considered a simple \emph{stop-feedback}  code that only uses feedback to inform the encoder of when to terminate of transmission.  A compelling example of this advantage can be seen for the BSC with capacity $1/2$ and target error probability $10^{-3}$. With variable-length coding and stop feedback, the average blocklength required to achieve $90\%$ of capacity is less than $200$, compared to at least $3100$ for the best fixed-blocklength code with noiseless feedback. 

In the non-asymptotic regime, Naghshvar \emph{et al.} asked the question of whether having two separate phases of operations and randomized encoding are necessary to achieve Burnashev's optimal error exponent. In \cite{Naghshvar2012}, they first presented a deterministic, one-phase coding scheme that achieves the optimal error exponent for any symmetric binary-input channels (including the BSC) with \emph{full, noiseless feedback}. The most appealing feature in their scheme is that at each time instant, the encoder only seeks a two-way partitioning of the message set such that the probability difference of the two subsets is ``small enough'' (see Sec. IV in \cite{Naghshvar2012}), and this is sufficient for their scheme to achieve both capacity and the optimal error exponent. Since the authors did not provide a name for their scheme, here we term their scheme as the \emph{small-enough-difference (SED) encoder}\footnote{We first coined this name in our conference paper \cite{Yang2020}.}. In a subsequent work \cite{Naghshvar2015}, Naghshvar \emph{et al.} applied the extrinsic Jensen-Shannon (EJS) divergence and submartingale synthesis technique to develop a non-asymptotic upper bound on the average blocklength for the SED encoder over symmetric binary-input channels with noiseless feedback. Recently, Guo \emph{et al.} \cite{Guo2021} developed an instantaneous SED code for the symmetric binary-input channels with feedback for real-time communication.

While Naghshvar \emph{et al.} obtained a non-asymptotic upper bound on average blocklength for their SED encoder, the resulting achievability bound 
falls beneath Polyanskiy's achievability bound for a system that only employs stop-feedback codes. In general, a system, such as the SED encoder, that employs full, noiseless, instantaneous feedback should achieve a rate much better than that of a stop-feedback code.  Thus, there is an opportunity to develop tighter lower bounds on the achievable rate of the SED encoder.  Furthermore, the SED encoder has not yet been extended to a general binary-input channel with feedback,  let alone a general multi-input DMC with feedback.

As a primary contribution, this paper extends Naghshvar \emph{et al.}'s SED encoder to the class of binary asymmetric channels (BACs) with feedback, including the BSC as a special case, and develops non-asymptotic upper bounds on average blocklength that are close to the actual performance of SED encoders. Unlike Naghshvar \emph{et al.}'s one-phase SED encoder, our SED encoder for a general BAC is a deterministic, two-phase encoder that performs a two-way partitioning of the message set such that \emph{the weighted probability difference} is small enough in the communication phase. In the confirmation phase, the encoder assigns the most distinguishable symbol exclusively to the most likely message. In particular, for the BSC with feedback, we develop a refined non-asymptotic upper bound on the average blocklength. Simulations demonstrate that both associated achievability bounds on rate exceed the stop-feedback achievability bound of Polyanskiy \emph{et al.} for the BSC with capacity $1/2$, which is expected since a system with full, noiseless feedback should perform better than one that is limited to stop feedback.

In our analysis, the technique for obtaining the bound for a general BAC involves a submartingale synthesis with optimal parameters. For the specific case of the BSC, the confirmation phase can be modeled as a Markov chain with possible fallbacks to the communication phase. This facilitates a decomposition of the random process concerning the transmitted message into two components:  a submartingale describing the first communication phase and a generalized Markov chain that describes the subsequent behavior (see Section \ref{subsec: proof of lemma 5}). This decomposition allows a separate upper bound to be computed for each of the two components.  The upper bound for the first component is obtained using a surrogate submartingale construction and a variant of Doob's optional stopping theorem. The upper bound for the second component is obtained using time of first-passage analysis on the generalized Markov chain. Finally, the sum of the two upper bounds yields an upper bound on the overall average blocklength that turns out to be tighter than the bound developed using purely submartingale synthesis when the crossover probability is small.

The remainder of this paper is organized as follows. In Section \ref{sec: problem setup}, we formulate the problem of variable-length coding over a BAC with noiseless feedback and review Naghshvar \emph{et al.}'s scheme for symmetric binary-input channels as well as some previous results. In Section \ref{sec: achievable rate BAC}, we present the SED encoder for a general BAC with noiseless feedback and a non-asymptotic upper bound on the corresponding average blocklength. In the case of the BSC with feedback, Section \ref{sec: achievable rate BSC} presents a new upper bound for the SED encoder developed by leveraging the submartingale synthesis and time of first passage analysis on Markov chains. Section \ref{sec: proofs} contains the proofs of the main results. In Section \ref{sec: numerical simulation}, we compare our bounds with the simulated performance of the SED encoder as well as some previously known results. In Section \ref{sec: reliability function}, we show that the proposed SED encoder achieves both capacity and the optimal error exponent of the BAC. Section \ref{sec: conclusion} concludes the paper.


\begin{figure}[t]
\centering
\includegraphics[width=0.45\textwidth]{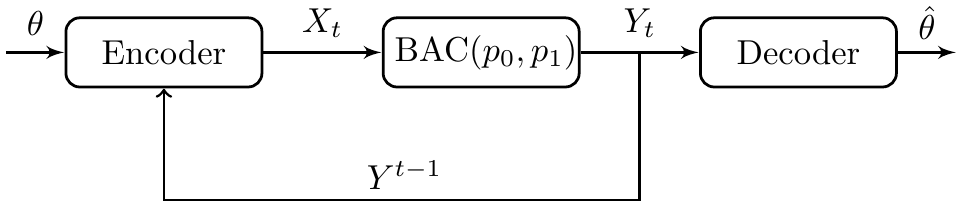}
\caption{A BAC$(p_0, p_1)$ with full, noiseless feedback link.}
\label{fig: system model}
\end{figure}

\section{Preliminaries}\label{sec: problem setup}

\subsection{Notation and Definitions}
Throughout the paper, $\log(\cdot), \ln(\cdot)$ denote the base-$2$ and the natural logarithms, respectively. $h(p) = -p\log(p)-(1-p)\log(1-p)$, $p\in[0, 1]$, denotes the binary entropy function. Let $P_Y, Q_Y$ be two distributions over a finite alphabet $\Y$, the \emph{Kullback-Leibler (KL) divergence} between $P_Y$ and $Q_Y$ is defined as $D(P_Y\| Q_Y)\triangleq \sum_{y\in\Y }P_Y(y)\log\frac{P_Y(y)}{Q_Y(y)}$ with the convention that $0\log\frac{0}{a} = 0$ and $b\log\frac{b}{0}=\infty$ for $a,b\in[0, 1]$ with $b\ne 0$. Let $[x]^+=\max\{0, x\}$. We denote the collection of all subsets of $\X$ by $2^\X$. 


\subsection{Problem Setup}
Consider the problem of variable-length coding over a BAC$(p_0, p_1)$ with noiseless feedback, as depicted in Fig. \ref{fig: system model}. The BAC consists of binary input and output alphabets, i.e., $\X = \Y = \Brace{0, 1}$, and two crossover probabilities, $p_0\triangleq P_{Y|X}(1|0)$ and $p_1 \triangleq P_{Y|X}(0|1)$. As noted in \cite{Moser2010}, it suffices to restrict our attention to the regularized case where $p_0\in(0, 1/2)$ and $p_0\le p_1\le 1-p_0$, as any other case can be transformed into this case by swapping either the input or output label. For ease of reference, we say a BAC$(p_0, p_1)$ is \emph{regularized} if the two crossover probabilities satisfy these conditions. If $p_0=p_1=p\in(0, 1/2)$, we simply write BSC$(p)$.

Let $C$ be the capacity of the BAC$(p_0, p_1)$ and let $(\pi_0^*, \pi_1^*)$ be the corresponding capacity-achieving input distribution. The following results will be useful in our proofs.

\begin{fact} \label{fact: 1}
  Consider a BAC$(p_0, p_1)$ with capacity achieving input distribution $(\pi_0^*, \pi_1^*)$. Then,
  \begin{align}
    C &= \frac{p_0h(p_1)}{1-p_0-p_1}-\frac{(1-p_1)h(p_0)}{1-p_0-p_1}+\log(1 + z),\label{eq: 1} \\
    \pi_0^* &= \frac{1 - p_1(1+z)}{(1-p_0-p_1)(1+z)},\label{eq: 2}\\
    \pi_1^* &= \frac{(1-p_0)(1+z)-1}{(1-p_0-p_1)(1+z)},\label{eq: 3}
  \end{align}
  where $z = 2^{\frac{h(p_0) - h(p_1)}{1-p_0-p_1}}$. Furthermore, if $p_0\in(0, 1/2)$ and $p_0\le p_1\le 1-p_0$, then $0 < \pi_1^*\le \pi_0^* < 1$.
\end{fact}
The proof of Fact \ref{fact: 1} is given in Appendix \ref{appendix: proof of fact 1}.

\begin{fact}[Theorem 4.5.1, \cite{Gallagerbook_1968}]\label{fact: 2}
  Consider a DMC with capacity-achieving input distribution $(\pi_0^*, \pi_1^*, \dots, \pi_{|\X|-1}^* )$. For each $k\in\Brace{0, 1, \dots, |\X|-1}$, if $\pi_k^* > 0$, then,
    \begin{align}
      D\Paren{P(Y|X=k)\Big\| \sum_{l=0}^{|\X|-1}\pi_l^*P(Y|X=l) } = C.
    \end{align}
\end{fact}

Let $C_1$ be the maximal KL divergence between two conditional output distributions, i.e.,
\begin{align}
  C_1 \triangleq \max_{x, x'\in\X} D\big(P(Y|X=x)\| P(Y|X = x')\big).
\end{align}
We also denote 
\begin{align}
  C_2 \triangleq \max_{y\in\Y}\log\frac{\max_{x\in \X}P_{Y|X}(y|x)}{\min_{x\in\X}P_{Y|X}(y|x) }.
\end{align}

\begin{fact}\label{fact: 3}
    For a regularized BAC$(p_0, p_1)$,
    \begin{align}
      C_1 &= D\Big(P(Y|X=1)\| P(Y|X = 0)\Big), \label{eq: 7}\\
      C_2 &= \log\frac{P_{Y|X}(1|1)}{P_{Y|X}(1|0)} = \log\frac{1-p_1}{p_0}. \label{eq: 8}
    \end{align}
\end{fact}
The proof of Fact \ref{fact: 3} is given in Appendix \ref{appendix: proof of fact 3}. 

For a regularized BAC, it always holds that $0 < C\le C_1 \le C_2 < \infty$. Later, we will see how these quantities are used in our result.

Let $\theta$ be the transmitted message uniformly drawn from the message set $\Omega = \Brace{1,2,\dots, M}$. The total transmission time (or the number of channel uses, or blocklength) $\tau$ is a random variable that is governed by a stopping rule that is a function of the observed channel outputs. Thanks to the full, noiseless feedback channel, the transmitter is also informed of the channel outputs and thus the stopping time.

The transmitter wishes to communicate $\theta$ to the receiver. To this end, it produces channel inputs $X_t$ for $t = 1, 2, \dots,\tau$ as a function of $\theta$ and past channel outputs $Y^{t-1} = (Y_1, Y_2, \dots, Y_{t-1})$, available to the transmitter through the noiseless feedback channel. Namely, 
\begin{align}
    X_t = e_t(\theta, Y^{t-1}), \quad t = 1, 2, \dots, \tau, \label{eq: 9}
\end{align}
for some encoding function $e_t:\Omega\times \Y^{t-1} \to \X$.

After observing $\tau$ channel outputs $Y_1, Y_2, \dots, Y_{\tau}$, the receiver makes a final estimate $\hat{\theta}$ of the transmitted message as a function of $Y^{\tau}$, i.e., 
\begin{align}
    \hat{\theta} = d(Y^{\tau}), \label{eq: 10}
\end{align}
for some decoding function $d: \Y^{\tau} \to \Omega$.

The probability of error of the transmission scheme is given by
\begin{align}
  P_e \triangleq \Prob\{\theta \ne \hat{\theta} \}.
\end{align}
For a fixed DMC (i.e., not necessarily restricted to BAC) and a given $\epsilon \in(0, 1/2)$, the general goal is to find encoding and decoding rules described in \eqref{eq: 9}, \eqref{eq: 10}, and a stopping time $\tau$ such that $P_e\le \epsilon$ and $\E[\tau]$ is minimized. Let $\E[\tau^*_{\epsilon}]$ be the minimum average blocklength. The achievable rate is defined as
\begin{align}
  R \triangleq \frac{\log M}{\E[\tau^*_{\epsilon}]}. \label{eq: achievable rate}
\end{align}
In \cite{Burnashev1976}, Burnashev, for the first time, derived the \emph{reliability function} $E(R)$ of variable-length coding over a fixed DMC for all rates $R < C$:
\begin{align}
  E(R) \triangleq \lim_{\epsilon\to0}\frac{-\log \epsilon}{\E[\tau^*_{\epsilon}]} = C_1\Paren{1 - \frac{R}{C}}.
\end{align}

\subsection{The SED Encoder of Naghshvar et al.}

In \cite{Naghshvar2012}, Naghshvar \emph{et al.} introduced a novel SED encoder for symmetric binary-input channels, which we now briefly describe as follows.

Let $\rho_i(t) \triangleq \Prob\Brace{\theta = i | Y^t}$, $t\ge 0$, be the posterior probability of $\theta = i$ given $Y^t$. Since $\theta$ is uniformly distributed before transmission, $\rho_i(0) = 1/M$ for all $i\in\Omega$. As noted in \cite{Naghshvar2015}, a sufficient statistic for estimating $\theta$ is the \emph{belief state} vector given by
\begin{align}
  \bm{\rho}(t) \triangleq [\rho_1(t), \rho_2(t), \dots, \rho_M(t)],\quad t = 0, 1, \dots, \tau.
\end{align}
According to the Bayes' rule, upon receiving $Y_t = y_t$, each $\rho_i(t)$, $i\in\Omega$, can be updated from $\bm{\rho}(t-1)$ by
\begin{align}
  \rho_i(t) = \frac{\rho_i(t-1)P_{Y|X}\big(y_t\,|\,e_t(i, Y^{t-1})\big) }{\sum_{j\in\Omega}\rho_j(t-1)P_{Y|X}\big(y_t\,|\,e_t(j, Y^{t-1})\big) }.
\end{align}
Thanks to the noiseless feedback, the transmitter will be informed of $y_t$ at time instant $t+1$ and thus can calculate the same $\bm{\rho}(t)$ before generating $X_{t+1}$.

\textbf{The SED encoder for BSC with feedback}: upon obtaining $\bm{\rho}(t)$ at time $t+1$, the encoder partitions the message set $\Omega$ into two subsets $S_0(t)$ and $S_1(t)$ such that
\begin{align}
  0\le \pi_0(t) - \pi_1(t) \le \min_{i\in S_0(t)}\rho_i(t), \label{eq: 16}
\end{align}
where $\pi_x(t) \triangleq \sum_{i\in S_x(t)}\rho_i(t)$, $x\in\{0, 1\}$. Once $S_0(t)$ and $S_1(t)$ are obtained, $X_{t+1} = 0$ if $\theta\in S_0(t)$ and $X_{t+1} = 1$ otherwise.

In the appendix of \cite{Naghshvar2012}, the authors demonstrated that \eqref{eq: 16} is sufficient to guarantee the achievability of both capacity and the optimal error exponent of the channel.

\subsection{The Decoder of Naghshvar et al. }
In \cite{Naghshvar2012} and \cite{Naghshvar2015}, Naghshvar \emph{et al.} considered the following possibly suboptimal stopping rule and decoder:
\begin{align}
  \tau &= \min\big\{ t: \max_{i\in\Omega}\rho_i(t)\ge 1-\epsilon \big\}, \label{eq: 17} \\
  \hat{\theta} &= \argmax_{i\in\Omega} \rho_i(\tau). \label{eq: 18}
\end{align}
Clearly, with the above scheme, the probability of error meets the desired constraint, i.e.,
\begin{align}
  P_e = \E[1 - \max_{i\in\Omega}\rho_i(\tau) ]\le \epsilon.
\end{align}
Let 
\begin{align}
U_i(t) \triangleq \log\frac{\rho_i(t)}{1 - \rho_i(t)} 
\end{align}
be the log-likelihood ratio of $\theta = i$ given $Y^t$. Equivalently, the stopping time in \eqref{eq: 17} can be written as
\begin{align}
  \tau = \min\left\{t: \max_{i\in\Omega} U_i(t) \ge \log\frac{1-\epsilon}{\epsilon} \right\}. \label{eq: 22}
\end{align}
In this paper, we consider the same possibly suboptimal stopping rule and decoder as described in \eqref{eq: 17} and \eqref{eq: 18}. Thus, the average blocklength $\E[\tau]$ only depends on the encoding scheme.


\subsection{Previous Results on Average Blocklength of VLF Codes}

In \cite{Naghshvar2015}, Naghshvar \emph{et al.} used the EJS divergence and submartingale synthesis technique to obtain a non-asymptotic upper bound on the average blocklength of the VLF code generated by their SED encoder for the symmetric binary-input channel with feedback.
\begin{theorem}[Remark 7, \cite{Naghshvar2015}]\label{theorem: 1}
  For a given $\epsilon\in(0, 1/2)$, the average blocklength of the VLF code generated by Naghshvar et al.'s SED encoder \eqref{eq: 16} and decoding rule \eqref{eq: 17} \eqref{eq: 18} for symmetric binary-input channels satisfies
  \begin{align}
    \E[\tau]\le \frac{\log M + \log\log\frac{M}{\epsilon}}{C} + \frac{\log\frac{1}{\epsilon}+1 }{C_1} + \frac{96\cdot 2^{2C_2}}{CC_1}. \label{eq: 20}
  \end{align}
\end{theorem}

The technique that underlies this result is a two-stage submartingale resulted from the SED encoding rule.
\begin{lemma}[\cite{Naghshvar2012}] \label{lemma: 1}
  Consider the SED encoder described in \eqref{eq: 16} for the BSC$(p)$, $p\in(0, 1/2)$, with feedback. If $\theta = i\in\Omega$, then $\Brace{U_i(t)}_{t=0}^{\infty}$ forms a submartingale with respect to the filtration $\F_t = \sigma\Brace{Y^t}$ satisfying
    \begin{subequations}
    \begin{align}
      \E[U_i(t+1)|\F_t,\theta=i] &\ge U_i(t) + C,\quad \text{if } U_i(t) < 0,\\
      \E[U_i(t+1)|\F_t,\theta=i] &= U_i(t) + C_1,\quad \text{if } U_i(t) \ge 0,\label{eq: 22b}\\
      |U_i(t+1) - U_i(t)| &\le C_2.
    \end{align}
    \end{subequations}
\end{lemma}
The proof of Lemma \ref{lemma: 1} can be found in \cite{Naghshvar2012}. We remark that the key step that links the SED encoder to the two-stage submartingale is the introduction and analysis of \emph{extrinsic probabilities}. This relation will be fully exploited in the proof of Lemma \ref{lemma: 3} (see Section \ref{subsec: proof of lemma 3}). The next step is to synthesize the two-stage submartingale in Lemma \ref{lemma: 1} into a single submartingale and then apply  Doob's optional stopping theorem. In \cite{Naghshvar2015}, with a sophisticated submartingale synthesis, Naghshvar \emph{et al.} obtained the following result.
\begin{lemma}[Lemma 8, \cite{Naghshvar2015}]\label{lemma: 2}
  Assume that the sequence $\Brace{\xi_t}_{t=0}^{\infty}$ forms a submartingale with respect to a filtration $\F_t$. Furthermore, assume there exist positive constants $K_1, K_2$ and $K_3$ such that
    \begin{subequations} 
    \begin{align}
      \E[\xi_{t+1}|\F_t] &\ge \xi_t + K_1, \quad \text{if } \xi_t < 0,\\
      \E[\xi_{t+1}|\F_t] &\ge \xi_t + K_2, \quad \text{if } \xi_t \ge 0,\\
      |\xi_{t+1} - \xi_t| &\le K_3,\quad \text{if } \max\Brace{\xi_{t+1}, \xi_t} \ge 0.
    \end{align}
    \end{subequations}
    Consider the stopping time $v = \min\Brace{t: \xi_t\ge B}$, $B > 0$. Then, we have the inequality,
    \begin{align}
      \E[v]\le \frac{B - \xi_0}{K_2} + \xi_0\indicator_{\Brace{\xi_0<0}} \Paren{\frac{1}{K_2} - \frac{1}{K_1}} + \frac{3K_3^2}{K_1K_2}.
    \end{align}
\end{lemma}
Observe that if $U_i(t)$ in Lemma \ref{lemma: 1} plays the role of $\xi_t$ in Lemma \ref{lemma: 2}, the sequence $\Brace{U_i(t)}_{t=0}^{\infty}$ meets the conditions in Lemma \ref{lemma: 2} by setting $K_1 = C$, $K_2 = C_1$ and $K_3 = C_2$. Thus, by setting $B = \log\frac{1-\epsilon}{\epsilon}$, the stopping rule in Lemma \ref{lemma: 2} coincides with that in \eqref{eq: 17} and we have the following corollary.
\begin{corollary}\label{corollary: 1}
  For a given $\epsilon\in(0, 1/2)$, the average blocklength of the VLF code generated by the SED encoder for BSC$(p)$, $p\in(0, 1/2)$, with feedback satisfies
  \begin{align}
  \E[\tau] \le \frac{\log M}{C} + \frac{\log\frac{1-\epsilon}{\epsilon}}{C_1} + \frac{3C_2^2}{CC_1}.\label{eq: 25}
  \end{align}
\end{corollary}
\begin{remark}
In \cite{Naghshvar2015}, Naghshvar et al. proved a two-stage submartingale similar to Lemma \ref{lemma: 1} by considering the average log-likelihood ratio $\tilde{U}(t)$ of the belief state $\bm{\rho}(t)$ rather than that of the transmitted message (see Appendix II in \cite{Naghshvar2015}). They showed that the average drift of $\tilde{U}(t)$ is characterized by the EJS divergence, which can be lower bounded by $C$ or $\tilde{\rho}C_1$ depending on the sign of $\tilde{U}(t)$, where $\tilde{\rho}\in(0, 1)$ is some constant. Combining their two-stage submartingale with Lemma \ref{lemma: 2}, they obtained Theorem \ref{theorem: 1}. However, a direct comparison of the third terms in  \eqref{eq: 20} and \eqref{eq: 25} immediately reveals that \eqref{eq: 25} is a significantly better upper bound on average blocklength.
\end{remark}

Next, we recall Polyanskiy's achievability result for an arbitrary DMC with feedback that utilizes a stop-feedback code.
\begin{theorem}[Theorem 3, \cite{Polyanskiy2011}]
  Consider a DMC with transition probability $P(y|x)$, $x\in\X$, $y\in \Y$. Fix a scalar $\gamma > 0$. Let $X^n$ and $\bar{X}^n$ be independent copies from the same process and let $Y^n$ be the output of the DMC when $X^n$ is the input. Define a sequence of information density functions
  \begin{align}
      \iota(a^n, b^n) \triangleq \log\frac{P_{Y^n|X^n}(a^n|b^n) }{P_{Y^n}(b^n)}
  \end{align}
  and a pair of hitting times 
  \begin{align}
      \psi &\triangleq \min\{n\ge0: \iota(X^n, Y^n)\ge \gamma \}, \\
      \bar{\psi} &\triangleq \min\{n\ge0: \iota(\bar{X}^n, Y^n)\ge \gamma \}.
  \end{align}
Then, for any $M$, there exists a VLF code satisfying
\begin{align}
    \E[\tau] &\le \E[\psi],\\
    P_e &\le (M - 1)\Prob\{\bar{\psi}\le \psi\}.
\end{align}
\end{theorem}



Finally, we recall Polyanskiy's converse bound for a VLF code with a non-vanishing error probability $\epsilon$.
\begin{theorem}[Theorems 4 and 6, \cite{Polyanskiy2011}]
  Consider a DMC with $0 < C \le C_1 < \infty$. Then any VLF code with $M$ codewords and target error probability $\epsilon$ satisfying $0 < \epsilon \le 1 - 1/M$ satisfies both
  \begin{align}
  &\E[\tau] \notag\\
  &\ge \sup_{0<\xi\le \frac{M-1}{M} }\Bigg[\frac1C\Paren{\log M {-} F_M(\xi) {-} \min\Big\{F_M(\epsilon), \frac{\epsilon}{\xi}\log M \Big\} } \notag\\
    &\phantom{=\,} + \Bigg[\frac{1-\epsilon}{C_1}\log\frac{\lambda_1\xi}{\epsilon(1-\xi)} - \frac{h(\epsilon)}{C_1} \Bigg]^+ \Bigg],
  \end{align}
 and
 \begin{align}
     \E[\tau]\ge \frac{(1-\epsilon)\log M - h(\epsilon)}{C},
 \end{align}
  where
  \begin{align}
    F_M(x) &\triangleq x\log(M-1) + h(x), x\in[0, 1], \\
    \lambda_1 &\triangleq \min_{y,x_1,x_2}\frac{P_{Y|X}(y|x_1)}{P_{Y|X}(y|x_2)}\in(0, 1).
  \end{align}
\end{theorem}

\section{Achievable Rates for BAC With Feedback}\label{sec: achievable rate BAC}

In this section, we introduce the SED encoder for a regularized BAC$(p_0, p_1)$ with noiseless feedback and develop a non-asymptotic upper bound on its average blocklength. Equivalently, this yields a lower bound on the achievable rate of the regularized BAC with feedback. 

For a general regularized BAC$(p_0, p_1)$, Naghshvar \emph{et al.}'s SED encoder no longer applies. As an extension, we propose the following deterministic, two-phase SED encoder for a regularized BAC$(p_0, p_1)$ with feedback.

\textbf{The SED encoder for regularized BAC$(p_0, p_1)$ with feedback}: upon obtaining $\bm{\rho}(t)$ at time $t+1$, let $\hat{i} = \argmax_{j\in\Omega}\rho_j(t)$. If $\rho_{\hat{i}}(t) < \pi_1^*$, the encoder partitions the message set $\Omega$ into two subsets $S_0(t)$ and $S_1(t)$ such that
\begin{align}
  -\min_{i\in S_1(t)}\rho_i(t)\le \frac{\pi_1^*}{\pi_0^*}\pi_0(t) - \pi_1(t) \le \frac{\pi_1^*}{\pi_0^*}\min_{i\in S_0(t)}\rho_i(t). \label{eq: SED encoder for BAC}
\end{align}
If $\rho_{\hat{i}}(t) \ge \pi_1^*$, the encoder exclusively assigns $S_1(t) = \{\hat{i}\}$ and $S_0(t) = \Omega\setminus\{\hat{i}\}$.
Once $S_0(t)$ and $S_1(t)$ are obtained, $X_{t+1} = 0$ if $\theta\in S_0(t)$ and $X_{t+1} = 1$ otherwise.

\begin{remark}
  First, we see that in the second case where $\rho_{\hat{i}}(t) \ge \pi_1^*$, the partition $S_1(t) = \{\hat{i}\}$, $S_0(t) = \Omega\setminus\{\hat{i}\}$ still meets \eqref{eq: SED encoder for BAC}. Second, if $p_0 = p_1$, then $\pi^*_0 = \pi^*_1 = 1/2$ and \eqref{eq: SED encoder for BAC} becomes
  \begin{align}
    -\min_{i\in S_1(t)}\rho_i(t)\le \pi_0(t) - \pi_1(t) \le \min_{i\in S_0(t)}\rho_i(t).
  \end{align}
  Clearly, this is a relaxation of \eqref{eq: 16} if the maximum posterior probability $\rho_{\hat{i}}(t) < 1/2$. If $\rho_{\hat{i}}(t) \ge 1/2$, \eqref{eq: 16} is met if and only if $S_0(t) = \{\hat{i} \}$ and $S_1(t) = \Omega\setminus\{\hat{i} \}$. In \cite{Naghshvar2012}, Naghshvar et al. showed that this assignment will yield \eqref{eq: 22b}. However, following their analysis, one can show that $S_1(t) = \{\hat{i} \}$ and $S_0(t) = \Omega\setminus\{\hat{i} \}$ also yield \eqref{eq: 22b}. Therefore, our SED encoder serves as a generalization of Naghshvar et al.'s  encoder.
\end{remark}

The motivation behind our SED encoder is that Lemma \ref{lemma: 1} now holds for the regularized BAC with feedback. For the sake of completeness, we state this result in a separate lemma.
\begin{lemma}\label{lemma: 3}
  Consider the SED encoder for a regularized BAC$(p_0, p_1)$ with feedback. If $\theta = i\in\Omega$, then $\Brace{U_i(t)}_{t=0}^{\infty}$ forms a submartingale with respect to the filtration $\F_t = \sigma\Brace{Y^t}$ satisfying
    \begin{subequations} \label{eq: 29}
    \begin{align}
      \E[U_i(t+1)|\F_t,\theta=i] &\ge U_i(t) + C,\quad \text{if } U_i(t) < 0, \label{eq: 29a}\\
      \E[U_i(t+1)|\F_t,\theta=i] &= U_i(t) + C_1,\quad \text{if } U_i(t) \ge 0,\label{eq: 29b}\\
      |U_i(t+1) - U_i(t)| &\le C_2. \label{eq: 29c}
    \end{align}
    \end{subequations}
\end{lemma}
\begin{IEEEproof}
 The proof fully exploits the properties of the extrinsic probabilities.  See Section \ref{subsec: proof of lemma 3} for more details.
\end{IEEEproof}

Since Lemma \ref{lemma: 2} is developed from a poor choice of parameters, here we perform a submartingale synthesis with optimized parameters to obtain the best possible upper bound on $\E[\tau]$ for a regularized BAC with feedback.
\begin{theorem}\label{theorem: BAC bound}
  For a given $\epsilon\in(0, 1/2)$, the average blocklength of the VLF code generated by the SED encoder for a regularized BAC$(p_0, p_1)$ with feedback satisfies
  \begin{align}
  \E[\tau]<& \frac{\log M}{C} {+} \frac{\log\frac{1-\epsilon}{\epsilon} + C_2 }{C_1} {+} C_2\Paren{\frac{1}{C} {-} \frac{1}{C_1}} \frac{1 - \frac{\epsilon}{1-\epsilon}2^{-C_2}}{1 - 2^{-C_2}}.
  \end{align}
\end{theorem}
\begin{IEEEproof}
    See Section \ref{subsec: proof of theorem 3}.
\end{IEEEproof}


\begin{algorithm}[t]
\caption{Original SED Encoding Algorithm}
\label{algorithm: 1}
\begin{algorithmic}[1]
  \Require $\max_{i\in \Omega}\rho_i < \pi_1^*$;
  \State $S_0\gets \{1,2,\dots, M\}$ and $S_1\gets \varnothing$;
  \State $\pi_0\gets1$, $\pi_1\gets0$, $\lambda\gets\pi_1^*/\pi_0^*$, $\delta\gets \lambda$, $\rho_{\min,0}\gets\min_{i\in S_0}\rho_i$, and $\rho_{\min, 1}\gets 0$;
  \While{$(\delta < -\rho_{\min,1})\ ||\ (\delta > \lambda\rho_{\min,0})$}
    \If{$\delta < -\rho_{\min,1}$}
      \State $j \gets \argmin_{i\in S_1}\rho_i$;
      \State $S_0\gets S_0\cup\{j\}$ and $S_1\gets S_1\setminus\{j\}$;
      \State $\pi_0\gets \pi_0 + \rho_j$ and $\pi_1\gets \pi_1 - \rho_j$;
    \EndIf
    \If{$\delta > \lambda\rho_{\min,0}$}
      \State $j \gets \argmin_{i\in S_0}\rho_i$;
      \State $S_0\gets S_0\setminus\{j\}$ and $S_1\gets S_1\cup\{j\}$;
      \State $\pi_0\gets \pi_0 - \rho_j$ and $\pi_1\gets \pi_1 + \rho_j$;
    \EndIf
    \State $\delta\gets \lambda\pi_0 - \pi_1$, $\rho_{\min,0}\gets \min_{i\in S_0}\rho_i$, $\rho_{\min,1}\gets \min_{i\in S_1}\rho_i$;
  \EndWhile
  \For{$i\gets 1, 2, \dots, M$}
    \State 
      $e_t(i, Y^{t-1}) = \begin{cases}
      0, & \text{if } i\in S_0\\
      1, & \text{if } i\in S_1
      \end{cases}$
  \EndFor
\end{algorithmic}
\end{algorithm}

\begin{algorithm}[t]
\caption{Greedy SED Encoding Algorithm}
\label{algorithm: 2}
\begin{algorithmic}[1]
  \Require $\max_{i\in \Omega}\rho_i < \pi_1^*$;
  \State $j_1\gets \argmax_{i\in\Omega}\rho_i$;
  \State $S_0 \gets\{j_1\}$ and $S_1\gets\varnothing$;
  \State $\pi_0 \gets \rho_{j_1}$, $\pi_1\gets 0$ and $\lambda\gets \pi_1^*/ \pi_0^*$;
  \For{$s\gets 2, 3, \dots, M$}
    \State $j_s \gets \argmax_{i\in\Omega\setminus\{j_1,\dots, j_{s-1}\}}\rho_i$;
    \If{$\pi_1 \ge \lambda\pi_0$}
      \State $S_0\gets S_0\cup\{j_s\}$;
      \State $\pi_0\gets \pi_0 + \rho_{j_s}$;
    \Else
      \State $S_1 \gets S_1\cup\{j_s\}$;
      \State $\pi_1\gets \pi_1 + \rho_{j_s}$;
    \EndIf
  \EndFor
  \For{$i\gets 1, 2, \dots, M$}
    \State 
      $e_t(i, Y^{t-1}) = \begin{cases}
      0, & \text{if } i\in S_0\\
      1, & \text{if } i\in S_1
      \end{cases}$
  \EndFor
\end{algorithmic}
\end{algorithm}

In the following, we show that the condition required by our SED encoder is always attainable at each time $t$. This is accomplished by solving a particular minimization problem.
\begin{theorem}\label{theorem: 4}
  For a regularized BAC with capacity-achieving input distribution $(\pi_0^*, \pi_1^*)$, let $\lambda \triangleq \pi_1^* / \pi_0^*\in(0, 1]$. For a given belief state vector $\bm{\rho} = [\rho_1, \rho_2, \dots, \rho_M]$ satisfying $\max_{i\in\Omega}\rho_i<\pi_1^*$, define the following objective function $f: 2^{\Omega} \to \R$:
  \begin{align}
    f(S)&\triangleq \lambda\big(\pi_1(S) - \lambda\pi_0(S)\big)\indicator_{\Brace{\pi_1(S)\ge \lambda\pi_0(S) }}\notag\\
      &\phantom{=\,} + \big(\lambda\pi_0(S) - \pi_1(S) \big)\indicator_{\Brace{\pi_1(S)<\lambda\pi_0(S) }},\label{eq: objective function}
  \end{align}
  where $\pi_0(S)\triangleq \sum_{i\in S}\rho_i$ and $\pi_1(S)\triangleq \sum_{i\in\Omega\setminus S}\rho_i$. Assume $S_0^*\subseteq\Omega$ minimizes \eqref{eq: objective function}. Then, the partition $(S_0^*, \Omega\setminus S_0^*)$ satisfies \eqref{eq: SED encoder for BAC}.
\end{theorem}
\begin{IEEEproof}
  See Section \ref{subsec: proof of theorem 4}.
\end{IEEEproof}

Theorem \ref{theorem: 4} implies that when the maximum posterior of $\bm{\rho}(t)$ does not exceed $\pi_1^*$, it is always possible to identify a two-way partition of $\Omega$ that satisfies \eqref{eq: SED encoder for BAC}. In fact, our proof already reveals such a partitioning algorithm as described in Algorithm \ref{algorithm: 1}. The algorithm is initialized with a partition of $\Omega$ that fails to meet \eqref{eq: SED encoder for BAC} and then successively constructs a new partition from the previous one to reduce $f(S)$. The termination condition is exactly given by \eqref{eq: SED encoder for BAC}. Theorem \ref{theorem: 4} guarantees that the termination will always be triggered at some point.

Finally, we present a greedy two-way partitioning algorithm as described in Algorithm \ref{algorithm: 2} that provably meets \eqref{eq: SED encoder for BAC}. We state this result in the following theorem.
\begin{theorem}\label{theorem: 5}
  Let $(\pi_0^*, \pi_1^*)$ be the capacity-achieving input distribution for a regularized BAC. Let $\bm{\rho} = [\rho_1, \rho_2, \dots, \rho_M]$ be the belief state vector for Algorithm \ref{algorithm: 2} satisfying $\max_{i\in\Omega}\rho_i < \pi_1^*$. Let $(S_0, S_1)$ be the partition of $\Omega$ generated by Algorithm \ref{algorithm: 2}. Then, $(S_0, S_1)$ meets the SED condition in \eqref{eq: SED encoder for BAC}.
\end{theorem}
\begin{IEEEproof}
  See Section \ref{subsec: proof of theorem 5}.
\end{IEEEproof}

\begin{remark}
  Both Algorithms \ref{algorithm: 1} and \ref{algorithm: 2} have complexity of order $O(M\log M)$, making them not suitable for practical implementation.
\end{remark}

\section{Achievable Rates for BSC With Feedback} \label{sec: achievable rate BSC}

In this section, we present our refined non-asymptotic upper bound on the average blocklength by adopting a SED encoder for the BSC$(p)$ with feedback. Both Naghshvar \emph{et al.}'s encoder and ours will yield the same result.

With the SED encoder and the BSC$(p)$, we obtain a refined upper bound on the average blocklength, as stated below.
\begin{theorem}\label{theorem: BSC bound}
  Let $q\triangleq 1 - p$. For a given $\epsilon\in(0, 1/2)$, the average blocklength of the VLF code generated by the SED encoder over the BSC$(p)$ with feedback, $p\in(0, 1/2)$,  satisfies
  \begin{align}
    \E[\tau]&< \frac{\log M}{C} + \frac{\log2q}{qC} + \frac{\log\frac{1-\epsilon}{\epsilon} + C_2 }{C_1} \notag\\
      &\phantom{=\,} + 2^{-C_2}C_2\left(\frac{1 + \frac{\log2q}{qC_2} }{C}-\frac{1}{C_1} \right)\frac{1 - \frac{\epsilon}{1-\epsilon}2^{-C_2}}{1 - 2^{-C_2}}.
  \end{align}
\end{theorem}
This result is a consequence of two supporting lemmas. To aid our discussion, let $q = 1-p$ and let us consider two stopping times for $\theta = i$ when $U_i(t)$ first crosses $0$ and $\log\frac{1-\epsilon}{\epsilon}$, respectively, 
\begin{align}
  \nu_i &\triangleq \min\{t: U_i(t)\ge 0\}, \\
  \tau_i &\triangleq \min\Brace{t: U_i(t)\ge \log\frac{1-\epsilon}{\epsilon}} . \label{eq: def tau_i}
\end{align}
Clearly, $\nu_i\le \tau_i$. $\nu_i$ and $\tau_i$ represent the stopping times when $\rho_i(t)$ first crosses $1/2$ and $1-\epsilon$, respectively. By Lemma \ref{lemma: 8}, both $\nu_i$ and $\tau_i$ are almost surely finite.

We are now in a position to introduce the two supporting lemmas. First, note that
\begin{align}
  \E[\tau] &= \frac{1}{M}\sum_{i=1}^M\E[\tau| \theta = i]\le \frac{1}{M}\sum_{i=1}^M\E[\tau_i| \theta = i], \label{eq: 36}
\end{align}
where the inequality follows since $\tau\le \tau_i$ for all $i\in\Omega$. Next, for $\E[\tau_i | \theta = i]$, it can be rewritten as
\begin{align}
  &\E[\tau_i | \theta = i] = \E[\nu_i | \theta = i] + \E[\tau_i-\nu_i|\theta = i]\\
    &= \E[\nu_i | \theta = i] + \E\big[\E[\tau_i-\nu_i|\theta = i, U_i(\nu_i) = u]|\theta=i\big]. \label{eq: 38}
\end{align}
The intuition behind this decomposition is that $\E[\nu_i | \theta = i]$ corresponds to the average blocklength in the \emph{first communication phase} (i.e., $U_i(t)$ from $\log(1/(M-1))$ to $0$), and $\E[\tau_i-\nu_i|\theta = i, U_i(\nu_i) = u]$ corresponds to the expected additional time spent in the confirmation phase with fallbacks to the communication phase. Here, $u$ represents the value at which $U_i(t)$ arrives when it crosses threshold $0$ for the first time.

Our next step is to develop upper bounds on $\E[\nu_i | \theta = i]$ and
$\E[\tau_i-\nu_i|\theta = i, U_i(\nu_i) = u]$ that are independent from $\theta = i$ and $U_i(\nu_i) = u$. Thus, summing up the two bounds will yield an upper bound on $\E[\tau_i|\theta = i]$, hence an upper bound on $\E[\tau]$ using \eqref{eq: 36}. We state our results in Lemmas \ref{lemma: first supporting lemma} and \ref{lemma: second supporting lemma}.

We remark that the technique for developing an upper bound on $\E[\nu_i | \theta = i]$ makes use of a \emph{surrogate submartingale}, thus allowing us to obtain a tighter constant term. In order to upper bound $\E[\tau_i-\nu_i|\theta = i, U_i(\nu_i) = u]$, we first observe that the behavior of $U_i(t)$ in the confirmation phase can be modeled as a Markov chain with a \emph{fallback self loop} on the initial state. This loop represents the probability that $U_i(t)$ first falls back to the communication phase and then returns to the confirmation phase. Next, we formulate the problem as solving a particular expected first-passage time on a Markov chain. The solution of this problem yields the desired upper bound. Detailed analysis can be found in the proof of each lemma.

\begin{lemma}\label{lemma: first supporting lemma}
  The average blocklength $\E[\nu_i]$ of a SED encoder over the BSC$(p)$, $p\in(0, 1/2)$, with feedback satisfies
  \begin{align}
    \E[\nu_i|\theta=i]< \frac{\log M}{C} + \frac{\log2q}{qC}.
  \end{align}
\end{lemma}

\begin{IEEEproof}
  See Section \ref{subsec: proof of lemma 4}.
\end{IEEEproof}

\begin{lemma}\label{lemma: second supporting lemma}
  Consider the SED encoder for the BSC$(p)$, $p\in(0, 1/2)$, with feedback. It holds that 
 \begin{align}
 &\E[\tau_i - \nu_i | \theta = i, U_i(\nu_i)=u] \notag\\
 &\le  \frac{\log\frac{1-\epsilon}{\epsilon} + C_2 }{C_1} {+} 2^{-C_2}C_2\left(\frac{1 + \frac{\log2q}{qC_2} }{C} {-} \frac{1}{C_1} \right)\frac{1 - \frac{\epsilon}{1-\epsilon}2^{-C_2}}{1 - 2^{-C_2}}.
 \end{align}
\end{lemma}

\begin{IEEEproof}
  See Section \ref{subsec: proof of lemma 5}.
\end{IEEEproof}


\section{Proofs}\label{sec: proofs}

In this section, we prove our main results.

\subsection{Proof of Lemma \ref{lemma: 3}}\label{subsec: proof of lemma 3}
Several steps in the proof of Lemma \ref{lemma: 3} are analogous to that in \cite{Naghshvar2012}, for instance, the introduction of the extrinsic probabilities. However, the distinction is that we will fully exploit the properties of the extrinsic probabilities that motivates our SED encoder for the BAC with feedback.

Let $\theta = i\in\Omega$ be fixed. For brevity, let $x_i$ be the input symbol for $\theta = i$ at time $t+1$. Let $\F_t=\sigma(Y^t)$ denote the filtration generated by $Y^t$. Thus, given $\F_t$ and $\theta = i$, $Y_{t+1}$ is distributed according to $P(Y|X = x_i)$. Hence,
\begin{align}
  &\E[U_{i}(t+1)-U_i(t)|\F_t, \theta = i ]\notag\\
  &= \sum_{y\in\Y}P_{Y|X}(y|x_i)\Paren{\log\frac{\rho_i(t+1)}{1-\rho_i(t+1)} - \log\frac{\rho_i(t)}{1-\rho_i(t)} }\notag\\
  &= \sum_{y\in\Y}P_{Y|X}(y|x_i)\notag\\
  &\phantom{=\,}\cdot\Paren{\log\frac{\frac{\rho_i(t)P_{Y|X}(y|x_i)}{\sum_{x\in\X}\pi_x(t)P_{Y|X}(y|x)} }{1-\frac{\rho_i(t)P_{Y|X}(y|x_i)}{\sum_{x\in\X}\pi_x(t)P_{Y|X}(y|x)} } - \log\frac{\rho_i(t)}{1-\rho_i(t)} }\\
  &=\sum_{y\in\Y}P_{Y|X}(y|x_i)\log\frac{P_{Y|X}(y|x_i)}{\sum_{x\in\X}\tilde{\pi}_{x,i}(t)P_{Y|X}(y|x) }\label{eq: extrinsic prob} \\
  &= D\big(P(Y|X=x_i)\| P(\tilde{Y}) \big), \label{eq: average step size}
\end{align}
where in \eqref{eq: extrinsic prob}, by letting $\bar{x} = 1-x$, we introduce the extrinsic probabilities defined by
\begin{align}
  \tilde{\pi}_{x_i, i}(t) &\triangleq \frac{\pi_{x_i}(t) - \rho_i(t) }{1 - \rho_i(t)}, \label{eq: extrinsic 1}\\
  \tilde{\pi}_{\bar{x}_i, i}(t) &\triangleq \frac{\pi_{\bar{x}_i}(t) }{1 - \rho_i(t)}, \label{eq: extrinsic 2}
\end{align}
and in \eqref{eq: average step size}, $\tilde{Y}$ is the output induced by the channel $P(Y|X)$ for an input $\tilde{X}$ distributed according to $(\tilde{\pi}_{0,i}(t), \tilde{\pi}_{1,i}(t))$.

Next, we prove the following key lemmas that connect our SED encoder to the two-stage submartingale in Lemma \ref{lemma: 3}.
\begin{lemma}\label{lemma: 4}
  The SED encoder for the regularized BAC$(p_0, p_1)$ with capacity-achieving input distribution $(\pi_0^*, \pi_1^*)$ satisfies $\tilde{\pi}_{x_i, i}(t)\le \pi_{x_i}^*$, where $x_i$ is the input symbol for $\theta = i$ at time $t+1$.
\end{lemma} 

\begin{IEEEproof}
  Let $\hat{i}=\argmax_{j\in\Omega}\rho_j(t)$. We distinguish two cases: $\rho_{\hat{i}}(t) < \pi_1^*$ and $\rho_{\hat{i}}(t) \ge \pi_1^*$.

  When $\rho_{\hat{i}}(t) < \pi_1^*$, we further discuss two subcases: $x_i = 0$ and $x_i = 1$. If $x_i = 0$, then $i\in S_0(t)$. Invoking the second inequality in \eqref{eq: SED encoder for BAC}, we have
  \begin{align}
    \tilde{\pi}_{0, i}(t) - \pi_0^* &= (\pi_0^* + \pi_1^*)\tilde{\pi}_{0, i}(t) - \pi_0^*\notag\\
      &= \pi_1^*\tilde{\pi}_{0,i}(t) - \pi_0^*(1 - \tilde{\pi}_{0,i}(t))\notag\\
      &= \pi_1^*\tilde{\pi}_{0,i}(t) - \pi_0^*\tilde{\pi}_{1,i}(t)\notag\\
      &= \frac{\pi_0^*}{1-\rho_i(t)}\Big(\frac{\pi_1^*}{\pi_0^*}\big(\pi_0(t)-\rho_i(t)\big) - \pi_1(t) \Big)\notag\\
      &\le \frac{\pi_0^*}{1-\rho_i(t)}\Big(\frac{\pi_1^*}{\pi_0^*}\big(\pi_0(t)-\min_{j\in S_0(t)}\rho_j(t)\big) - \pi_1(t) \Big)\notag\\
      &\le 0.
  \end{align}
  If $x_i = 1$, then $i\in S_1(t)$. In a similar fashion,
    \begin{align}
    \tilde{\pi}_{1, i}(t) - \pi_1^* &= \frac{\pi_0^*}{1-\rho_i(t)}\Big(\big(\pi_1(t)-\rho_i(t)\big) - \frac{\pi_1^*}{\pi_0^*}\pi_0(t) \Big)\notag\\
      &\le \frac{\pi_0^*}{1-\rho_i(t)}\Big(\big(\pi_1(t)-\min_{j\in S_1(t)}\rho_j(t)\big) - \frac{\pi_1^*}{\pi_0^*}\pi_0(t) \Big)\notag\\
      &\le0.
    \end{align}
    Therefore, Lemma \ref{lemma: 4} holds for $\rho_{\hat{i}}(t) < \pi_1^*$.

When $\rho_{\hat{i}}(t) \ge \pi_1^*$, by the encoding rule, $S_1(t) = \{\hat{i}\}$ and $S_0(t) = \Omega\setminus\{\hat{i}\}$. If $\hat{i} = i$, then $S_1(t) = \{i\}$ and $S_0(t) = \Omega\setminus\{i \}$. Thus, $\tilde{\pi}_{1,i}(t) = 0 < \pi_1^*$. If $\hat{i} \ne i$, then $i\in S_0(t)$. Since $\pi_1(t) = \rho_{\hat{i}}(t) \ge \pi_1^*$, it follows that $\pi_0(t)\le \pi_0^*$. Combining with the fact that $\tilde{\pi}_{0, i}(t) \le \pi_0(t)$, we conclude that $\tilde{\pi}_{0, i}(t) \le \pi_0^*$. Therefore, Lemma \ref{lemma: 4} also holds in this case.

    Summarizing the above two cases, we conclude that Lemma \ref{lemma: 4} holds in general.
\end{IEEEproof}

Next, we borrow a useful lemma on the KL divergence proved in \cite{Naghshvar2015}.

\begin{lemma}[Lemma 1, \cite{Naghshvar2015}]\label{lemma: 5}
  For any two distributions $P$ and $Q$ on a set $\Y$ and $\alpha\in[0, 1]$, $D(P\| \alpha P + (1-\alpha)Q)$ is decreasing in $\alpha$.
\end{lemma}

As an application of Lemma \ref{lemma: 5}, let $P = P(Y|X=x_i)$, $Q = P(Y|X = \bar{x}_i)$ and $\alpha = \tilde{\pi}_{x_i,i}(t)$. \eqref{eq: average step size} can be lower bounded by
\begin{align}
    D\big(P(Y|X=x_i)\| P(\tilde{Y}) \big) &= D\big(P\| \alpha P + (1-\alpha)Q \big)\\
    &\ge D\big(P\| \pi_{x_i}^* P + \pi_{\bar{x}_i}^* Q \big)\label{eq: 38}\\
    &= C,\label{eq: 39}
\end{align}
where \eqref{eq: 38} follows from Lemma \ref{lemma: 4} and \eqref{eq: 39} follows from Fact \ref{fact: 2}. Therefore, with the SED encoder, it always holds that
\begin{align}
  \E[U_i(t+1)|\F_t,\theta=i] &\ge U_i(t) + C.
\end{align}
As a result, \eqref{eq: 29a} is proved.

In particular, if $U_i(t)\ge0$, this is equivalent to $\rho_i(t)\ge 1/2$ and thus $i$ is the index with the maximum posterior. Using Fact \ref{fact: 1} that $\pi_1^*\le 1/2$, it follows that $\max_{j\in\Omega}\rho_j(t) = \rho_i(t) \ge \pi_1^*$. Thus, the SED encoder will exclusively assign $S_1(t) = \{i\}$ and $S_0(t) = \Omega\setminus\{i\}$, resulting in $\tilde{\pi}_{1, i}(t) = 0$ and
\begin{align}
  D\big(P(Y|X=x_i)\| P(\tilde{Y}) \big) &= D\big(P(Y|X=1)\| P(Y|X=0) \big)\notag\\
    &= C_1, \label{eq: 41}
\end{align}
where \eqref{eq: 41} follows from Fact \ref{fact: 3}. Therefore, \eqref{eq: 29b} is proved. We also remark that for $Y_{t+1} = y$,
\begin{align}
  U_i(t+1) = U_i(t) + \log\frac{P_{Y|X}(y|1)}{P_{Y|X}(y|0)},\quad \text{if } U_i(t)\ge0. \label{eq: instantaneous update}
\end{align}
Hence, $C_1$ can be thought of as the average drift of $U_i(t)$ for $U_i(t)\ge 0$.

To prove \eqref{eq: 29c}, we note that when $Y_{t+1} = y$,
\begin{align}
  &|U_i(t+1) - U_i(t)|\notag\\
  &= \left|\log\frac{\rho_i(t+1)}{1-\rho_i(t+1)} - \log\frac{\rho_i(t)}{1 - \rho_i(t)}  \right|\notag\\
  &= \left|\log\left(\frac{\rho_i(t)P_{Y|X}(y\,|\,e_{t+1}(i, Y^t)) }{\sum_{j\ne i}\rho_j(t)P_{Y|X}(y\,|\, e_{t+1}(j, Y^t)) }\cdot\frac{1-\rho_i(t)}{\rho_i(t)}\right)  \right|\notag\\
  &= \left|\log\frac{P_{Y|X}(y\,|\,e_{t+1}(i, Y^t)) }{\sum_{j\ne i}\frac{\rho_j(t)}{1-\rho_i(t)} P_{Y|X}(y\,|\, e_{t+1}(j, Y^t)) }  \right|\\
  &\le \log\frac{\max_{x\in\X}P_{Y|X}(y|x) }{\min_{x\in\X}P_{Y|X}(y|x)}.
\end{align}
Hence, we have
\begin{align}
  |U_i(t+1) - U_i(t)|\le \max_{y\in\Y}\log\frac{\max_{x\in\X}P_{Y|X}(y|x) }{\min_{x\in\X}P_{Y|X}(y|x)},
\end{align}
which completes the proof of \eqref{eq: 29c}.

\subsection{Proof of Theorem \ref{theorem: BAC bound}}\label{subsec: proof of theorem 3}

The proof of Theorem \ref{theorem: BAC bound} involves a submartingale synthesis with optimized parameters and a variant of Doob's optional stopping theorem. Throughout the proof, we fix $\theta = i\in\Omega$ to avoid writing the conditioning $\theta = i$ unless otherwise specified.

Let the sequence $\{U_i(t)\}_{t=0}^\infty$ be the two-stage submartingale defined in \eqref{eq: 29} with respect to filtration $\{\F_t\}_{t=0}^{\infty}$ as a result of the SED encoding over a regularized BAC$(p_0, p_1)$. Let us consider a sequence $\{\eta(t)\}_{t=0}^{\infty}$ defined as
\begin{align}
  \eta(t) = \begin{cases}
      -A + \frac{U_i(t)}{C} - t,\quad \text{if } U_i(t) < 0,\\
      -Ae^{-sU_i(t)} + \frac{U_i(t)}{C_1} -t,\quad \text{if } U_i(t)\ge 0,
  \end{cases}\label{eq: eta_t}
\end{align}
where $s > 0$ and $A > 0$ are two constants. For our purposes, we require that $s$ and $A$ meet the following two equations.
\begin{align}
    &A(1 - e^{-sC_2}) - C_2\Paren{\frac{1}{C} - \frac{1}{C_1}} = 0, \label{eq: first eq}\\
    &p_1 e^{-s\log\frac{p_1}{1-p_0}} + (1-p_1)e^{-s\log\frac{1-p_1}{p_0}} = 1. \label{eq: second eq}
\end{align}
The motivation behind these equations is to select the best parameters that make $\{\eta(t)\}_{t=0}^{\infty}$ a submartingale. This will become clearer as our proof proceeds. Solving \eqref{eq: first eq} and \eqref{eq: second eq} for $s$ and $A$ yields
\begin{align}
  s &= \ln 2, \label{eq: param s}\\
  A &= \frac{C_2}{1 - 2^{-C_2}}\Paren{\frac{1}{C} - \frac{1}{C_1}}. \label{eq: param A}
\end{align}

\begin{lemma}\label{lemma: 6}
  The sequence $\{\eta(t)\}_{t=0}^{\infty}$ with parameters $s$ and $A$ satisfying \eqref{eq: first eq} and \eqref{eq: second eq} forms a submartingale with respect to the filtration $\{\F_t\}_{t=0}^{\infty}$.
\end{lemma}
\begin{IEEEproof}
  We will show that $\E[\eta(t+1)|\F_t]\ge \eta(t)$. There are two cases.

  \textit{Case 1 $(U_i(t) < 0)$}: there are two subcases. If $U_i(t+1)\ge 0$, then from \eqref{eq: 29c}, $U_i(t+1) < C_2$. Consider the function
  \begin{align}
    f(x) \triangleq A - Ae^{-sx} - \Paren{\frac{1}{C}-\frac{1}{C_1} }x , \label{eq: f function}
  \end{align}
  where $s$ and $A$ satisfy equations \eqref{eq: first eq} and \eqref{eq: second eq}. Since $f(0) = 0$, $f(C_2) = 0$ due to \eqref{eq: first eq}, and $f(x)$ is a concave function, it follows that $f(x) > 0$ for $x\in(0, C_2)$. Let $U_i(t+1)$ play the role of $x$. Using $f(U_i(t+1))>0$, we obtain
  \begin{align}
    \eta(t+1) &= -Ae^{-sU_i(t+1)} + \frac{U_i(t+1)}{C_1} - (t+1)\notag\\
     &> -A + \frac{U_i(t+1)}{C} - (t+1).
  \end{align}
  If $U_i(t+1) < 0$, then
  \begin{align}
    \eta(t+1) &= -A + \frac{U_i(t+1)}{C} - (t+1).
  \end{align}
  Hence, regardless of the sign of $U_i(t+1)$, it holds that
  \begin{align}
    \E[\eta(t+1)|\F_t] &\ge \E\Big[-A + \frac{U_i(t+1)}{C} - (t+1) \Big|\F_t \Big]\\
    &\ge -A + \frac{U_i(t) + C}{C} - (t+1)\label{eq: 55}\\
    &= \eta(t),
  \end{align}
  where \eqref{eq: 55} follows from \eqref{eq: 29a}.

  \textit{Case 2 $(U_i(t)\ge 0)$}: there are two subcases. If $U_i(t+1) < 0$, using $f(x)$ defined in \eqref{eq: f function}, $f(U_i(t+1)) < 0$. Therefore,
    \begin{align}
      \eta(t+1) &= -A + \frac{U_i(t+1)}{C} - (t+1)\\
        &\ge -Ae^{-sU_i(t+1)} + \frac{U_i(t+1)}{C_1} - (t+1).
    \end{align}
    If $U_i(t+1)\ge 0$, then
    \begin{align}
      \eta(t+1) &= -Ae^{-sU_i(t+1)} + \frac{U_i(t+1)}{C_1} - (t+1).
    \end{align}
    Hence, regardless of the sign of $U_i(t+1)$, it holds that
    \begin{align}
      &\E[\eta(t+1)|\F_t]\notag\\
      &\ge \E\Big[-Ae^{-sU_i(t+1)} + \frac{U_i(t+1)}{C_1} - (t+1) \Big|\F_t \Big]\\
      &= -A\E[e^{-sU_i(t+1)}|\F_t] + \frac{\E[U_i(t+1)|\F_t]}{C_1} - (t+1)\\
      &= -A\Big(p_1e^{-s\log\frac{p_1}{1-p_0}}+(1-p_1)e^{-s\log\frac{1-p_1}{p_0}} \Big)e^{-sU_i(t)} \notag\\
        &\phantom{=\,} + \frac{U_i(t)}{C_1} - t \label{eq: 62} \\
      &= \eta(t), \label{eq: 63}
    \end{align}
    where \eqref{eq: 62} follows from \eqref{eq: 29b} and \eqref{eq: instantaneous update}, and \eqref{eq: 63} follows from \eqref{eq: second eq}.

    Summarizing the above two cases, we conclude that $\E[\eta(t+1)|\F_t]\ge \eta(t)$.
\end{IEEEproof}


\begin{lemma}\label{lemma: 7}
    The sequence $\{\eta(t)\}_{t=0}^{\infty}$ with parameters $s$ and $A$ satisfying \eqref{eq: first eq} and \eqref{eq: second eq} has the property that the difference between $\eta(t+1)$ and $\eta(t)$ is absolutely bounded, i.e., 
    \begin{align}
        |\eta(t+1) - \eta(t)| \le A + \frac{2C_2}{C} + 1.
    \end{align}
\end{lemma}
\begin{IEEEproof}
  We distinguish four cases.

  \textit{Case 1}: $U_i(t) < 0$ and $U_i(t+1) < 0$. In this case,
    \begin{align}
      |\eta(t+1) - \eta(t)| &= \left|\frac{U_i(t+1)-U_i(t)}{C}-1 \right| \notag\\
      &\le \frac{|U_i(t+1)-U_i(t)|}{C} + 1 \notag\\
      &\le \frac{C_2}{C} + 1 \label{eq: UB1}
    \end{align}
  \textit{Case 2}: $U_i(t) < 0$ and $U_i(t+1) \ge 0$. In this case, $U_i(t+1)\le C_2$ by \eqref{eq: 29c}, and
  \begin{align}
    &|\eta(t+1) - \eta(t)|\notag\\
  &= \left|A(1 - e^{-sU_i(t+1)}) + \frac{U_i(t+1)}{C_1} - \frac{U_i(t)}{C} -1 \right| \notag\\
  &\le A(1 - e^{-sC_2}) + \frac{C_2}{C} + 1. \label{eq: UB2}
  \end{align}
  \textit{Case 3}: $U_i(t) \ge 0$ and $U_i(t+1) < 0$. In this case, $U_i(t)\le C_2$ by \eqref{eq: 29c}, and
  \begin{align}
    &|\eta(t+1) - \eta(t)|\notag\\
  &= \left|A(e^{-sU_i(t)}-1) + \frac{U_i(t+1)}{C} - \frac{U_i(t)}{C_1} -1 \right| \notag\\
  &\le A|1 - e^{-sU_i(t)}| \notag\\
    &\phantom{=\,} + \left|\frac{U_i(t+1)-U_i(t)}{C} + \left(\frac{1}{C}-\frac{1}{C_1}\right)U_i(t) \right| + 1\notag\\
  &\le A(1 - e^{-sC_2}) + \frac{C_2}{C} + \left(\frac{1}{C}-\frac{1}{C_1} \right)C_2 + 1. \label{eq: UB3}
  \end{align}
  \textit{Case 4}: $U_i(t)\ge0$ and $U_i(t+1)\ge0$. In this case,
  \begin{align}
    &|\eta(t+1) - \eta(t)|\notag\\
  &= \left|-A\big(e^{-sU_i(t+1)} - e^{-sU_i(t)}\big) + \frac{U_i(t+1)-U_i(t)}{C_1} -1 \right|\notag\\
  &\le A\big|e^{-sU_i(t+1)} - e^{-sU_i(t)}\big| + \frac{|U_i(t+1)-U_i(t)|}{C_1} + 1 \notag\\
  &\le A(1 - e^{-sC_2}) + \frac{C_2}{C_1} + 1, \label{eq: UB4}
  \end{align}
  where \eqref{eq: UB4} follows from the inequality $|e^{-sy} - e^{-sx}|\le 1 - e^{-s|y-x|}$ for $s\ge 0$, $x\ge 0$ and $y\ge 0$.

  Note that the upper bounds in \eqref{eq: UB1}, \eqref{eq: UB2}, \eqref{eq: UB3} and \eqref{eq: UB4} are no greater than $A + \frac{2C_2}{C} + 1$. The proof is completed.
\end{IEEEproof}

\begin{lemma}\label{lemma: 8}
  Let $\{U(t)\}_{t=0}^{\infty}$ be the submartingale in \eqref{eq: 29} with respect to filtration $\{\F_t\}_{t=0}^\infty$. Consider the stopping time $T \triangleq \min\{t:  U(t)\ge \zeta\}$, where $\zeta > 0$ is some constant. Then, $\Prob\{T < \infty\} = 1$. Namely, $T$ is a.s. finite.
\end{lemma}
\begin{IEEEproof}
  We first recall Azuma's inequality for a general submartingale $\{\xi(t)\}_{t=0}^\infty$: If $\{\xi(t)\}_{t=0}^\infty$ is a submartingale that satisfies $|\xi(t+1) - \xi(t)|\le K$ for all $t\ge 0$, then for a given $\sigma > 0$,
  \begin{align}
    \Prob\{\xi(t) - \xi(0) \le -\sigma \} \le \exp\left(\frac{-\sigma^2}{2tK^2} \right).
  \end{align}
  Let us consider $\xi(t) \triangleq \frac{U(t)}{C} - t$. We show that $\{\xi(t)\}_{t=0}^\infty$ is also a submartingale with respect to filtration $\{\F_t\}_{t=0}^\infty$. Specifically, if $U(t) < 0$, then
  \begin{align}
    \E[\xi(t+1) | \F_t] &= \frac{\E[U(t+1)|\F_t]}{C} - (t+1)\\
      &\ge \frac{U(t) + C}{C} - (t+1)\\
      &= \xi(t).
  \end{align}
  If $U(t)\ge 0$, using the fact that $C_1 \ge C$, we can also show that $\E[\xi(t+1) | \F_t]\ge \xi(t)$. Hence, $\{\xi(t)\}_{t=0}^\infty$ is a submartingale with respect to filtration $\{\F_t\}_{t=0}^\infty$. Furthermore, for any $t\ge 0$,
  \begin{align}
    |\xi(t+1) - \xi(t)| = \left|\frac{U(t+1)-U(t)}{C}-1 \right|\le \frac{C_2}{C}+1.
  \end{align}
  Let $K = \frac{C_2}{C}+1$ for shorthand notation. Thus, appealing to Azuma's inequality,
  \begin{align}
    &\Prob\Brace{U(t)\le (t-\sigma)C + U(0) }\notag\\
    &=\Prob\Brace{\frac{U(t)}{C} - t - \frac{U(0)}{C} \le -\sigma }\\
    &=\Prob\Brace{\xi(t) - \xi(0)\le -\sigma }\\
    &\le \exp\Paren{\frac{-\sigma^2}{2tK^2 } }.
  \end{align}
  Equating $\zeta = (t - \sigma)C + U(0)$ yields $\sigma = t - \frac{\zeta - U(0)}{C}$, $t>\frac{\zeta - U(0)}{C}$. Hence,
  \begin{align}
  \Prob\Brace{U(t)\le \zeta }&\le \exp\Paren{\frac{-(t - \frac{\zeta - U(0)}{C})^2}{2tK^2} }\\
    &= \exp\Paren{-\frac{t}{2K^2} + O(t^{-1}) }.
  \end{align}
  It follows that
  \begin{align}
    \lim_{t\to\infty}\Prob\Brace{U(t)\le \zeta} \le \lim_{t\to\infty} \exp\Paren{-\frac{t}{2K^2} + O(t^{-1}) } = 0.
  \end{align}
  This implies that
  \begin{align}
    \Prob\Brace{T = \infty} &= \lim_{t\to\infty}\Prob\Paren{\bigcap_{k=1}^t\Brace{U(k) < \zeta } }\\
        &\le \lim_{t\to\infty}\Prob\Brace{U(t)\le \zeta}\\
        &=0.
  \end{align}
  Namely, $\Prob\Brace{T < \infty}  =1$.
\end{IEEEproof}

Finally, we follow \cite{Burnashev1975} to prove a variant of Doob's optional stopping theorem which will be useful in proving the main result.
\begin{lemma}[Variant of Doob's Optional Stopping Theorem]\label{lemma: variant of Doob}
  Let $\Brace{U(t)}_{t=0}^\infty$ be a submartingale with respect to filtration $\Brace{\F_t}_{t=0}^\infty$ satisfying $|U(t+1) - U(t)|\le K$ for some positive constant $K$. Let $T = \min\{t: U(t)\ge \zeta\}$, $\zeta>0$ be a stopping time and assume that $T$ is a.s. finite. Then, 
  \begin{align}
      U(0) \le \E[U(T)].
  \end{align}
\end{lemma}

\begin{IEEEproof}
  Let $t\wedge T \triangleq \min\{t, T\}$. From the martingale theory \cite{Williams_1991}, the stopped process $\{U(t\wedge T)\}_{t=0}^\infty$ is also a submartingale. Thus, we obtain
  \begin{align}
      U(0) &\le \E[U(t\wedge T)] \label{eq: step1} \\
        &\le \lim_{t\to \infty} \E[U(t\wedge T)] \label{eq: step2}\\
        &\le \E[\lim_{t\to \infty} U(t\wedge T)] \label{eq: step3}\\
        &= \E[U(T)]. \notag
  \end{align}
In the above, 
\begin{itemize}
    \item \eqref{eq: step1} follows from applying Doob's optional stopping theorem \cite{Williams_1991} to the stopped process $\{U(t\wedge T)\}_{t=0}^\infty$.
    \item \eqref{eq: step2} follows from that $\E[U(t\wedge T)]\le \E[U((t+1)\wedge T)]$ for submartingales. This can be seen by noting that
    \begin{align*}
        &\E[U((t+1)\wedge T)] - \E[U(t\wedge T)] \notag\\
        &= \E[(U(t+1) - U(t))\indicator_{\Brace{T\ge t+1}} ] + \E[0\cdot\indicator_{\Brace{T\le n}}]\notag\\
        &=\E\big[\E[(U(t+1) - U(t))\indicator_{\Brace{T\ge t+1}} | \F_t ]\big]\\
        &= \E\big[\indicator_{\Brace{T\ge t+1}}\E[(U(t+1) - U(t)) | \F_t ]\big]\\
        &\ge 0,
    \end{align*}
    where the last step follows from submartingale property $\E[U(t+1)|\F_t]\ge U(t)$.
    \item \eqref{eq: step3} follows from the fact that $U(t\wedge T)$ is uniformly bounded above, the assumption that $T$ is a.s. finite, and the reverse Fatou's lemma.
\end{itemize}
This concludes the proof of Lemma \ref{lemma: variant of Doob}.
\end{IEEEproof}

Let us consider the stopping time for each $j\in\Omega$:
\begin{align}
  \tau_j \triangleq \min\left\{t: U_j(t)\ge \log\frac{1-\epsilon}{\epsilon} \right\},\quad j\in\Omega.  \label{eq: tau_i}
\end{align}
Lemmas \ref{lemma: 6}, \ref{lemma: 7} and \ref{lemma: 8} indicate that the submartingale $\{\eta(t)\}_{t=0}^\infty$ in \eqref{eq: eta_t} with parameters $s$ and $A$ given by \eqref{eq: param s} and \eqref{eq: param A} and the stopping time $\tau_i$ in \eqref{eq: tau_i} meet the conditions in Lemma \ref{lemma: variant of Doob}. Hence, by Lemma \ref{lemma: variant of Doob},

\begin{align}
    &\eta(0) \le \E[\eta(\tau_i)|\theta = i]\notag\\
      &= \E\Br{-Ae^{-sU_i(\tau_i)} + \frac{U_i(\tau_i)}{C_1} - \tau \Big|\theta = i }\\
      &\le -Ae^{-s(\log\frac{1-\epsilon}{\epsilon} + C_2)} + \frac{\log\frac{1-\epsilon}{\epsilon}+C_2 }{C_1} - \E[\tau_i|\theta=i], \label{eq: 85}
\end{align}
where \eqref{eq: 85} follows since
\begin{align}
  \E[U_i(\tau)] &= \E[U_i(\tau_i) - U_i(\tau_i-1)] + \E[U_i(\tau_i-1)]\\
    &< C_2 + \log\frac{1-\epsilon}{\epsilon}.
\end{align}
Rewriting \eqref{eq: 85} and substituting $s$ and $A$ with \eqref{eq: param s} and \eqref{eq: param A} respectively yield
\begin{align}
  &\E[\tau_i|\theta = i]\le -Ae^{-s(\log\frac{1-\epsilon}{\epsilon} + C_2)} + \frac{\log\frac{1-\epsilon}{\epsilon}+C_2 }{C_1} - \eta(0)\notag\\
    &= -Ae^{-s(\log\frac{1-\epsilon}{\epsilon} + C_2)} + \frac{\log\frac{1-\epsilon}{\epsilon}+C_2 }{C_1} + A - \frac{U_i(0)}{C}\notag\\
    &< \frac{\log M}{C} + \frac{\log\frac{1-\epsilon}{\epsilon} + C_2 }{C_1} + C_2\Paren{\frac{1}{C} {-} \frac{1}{C_1}} \frac{1 - \frac{\epsilon}{1-\epsilon}2^{-C_2}}{1 - 2^{-C_2}}. \label{eq: 90}
\end{align}
Finally,
\begin{align}
  \E[\tau] &= \frac{1}{M}\sum_{j=1}^M \E[\tau | \theta = j]\le \frac{1}{M}\sum_{i=1}^M \E[\tau_j | \theta = j], \label{eq: 91}
\end{align}
where the last inequality follows since $\tau\le \tau_j$ for all $j\in\Omega$. Since the upper bound in \eqref{eq: 90} holds for any $\E[\tau_j|\theta = j]$. Substituting the bound in \eqref{eq: 91} completes the proof of Theorem \ref{theorem: BAC bound}.

\subsection{Proof of Theorem \ref{theorem: 4}}\label{subsec: proof of theorem 4}

We prove Theorem \ref{theorem: 4} by contradiction. Let $S_0^*\subseteq \Omega$ be an optimal subset of $\Omega$ that minimizes $f(S)$ in \eqref{eq: objective function}. If the partition $(S_0^*, \Omega\setminus S_0^*)$ does not meet \eqref{eq: SED encoder for BAC}, one can construct another subset $S_0'\subseteq\Omega$ from $S_0^*$ such that $f(S_0') < f(S_0^*)$, thus contradicting the assumption that $S_0^*$ minimizes $f(S)$.

Assume that the partition $(S_0^*, \Omega\setminus S_0^*)$ does not meet \eqref{eq: SED encoder for BAC}, there are two cases.

\textit{Case 1}: the partition $(S_0^*, \Omega\setminus S_0^*)$ satisfies $\lambda\pi_0(S_0^*) - \pi_1(S_0^*)< - \min_{i\in\Omega\setminus S_0^*}\rho_i $. Let $i^* = \argmin_{i\in\Omega\setminus S_0^*}\rho_i$. Then,
\begin{align}
  f(S_0^*) &= \lambda\big(\pi_1(S_0^*) - \lambda\pi_0(S_0^*)\big) > \lambda \rho_{i^*}. \label{eq: 94}
\end{align}
Consider a new subset $S_0' \triangleq S_0^*\cup\{i^* \}$. Next, we show that $f(S_0') < f(S_0^*)$. There are two subcases. If $\pi_1(S_0')\ge \lambda\pi_0(S_0')$, then
\begin{align}
    f(S_0') &= \lambda\big(\pi_1(S_0') - \lambda\pi_0(S_0') \big) \notag\\
      &= \lambda\big(\pi_1(S_0^*) - \rho_{i^*} - \lambda\pi_0(S_0^*) - \lambda\rho_{i^*} \big)\notag\\
      &< \lambda\big(\pi_1(S_0^*) - \lambda\pi_0(S_0^*) \big)\label{eq: 95}\\
      &= f(S_0^*),\notag
\end{align}
where \eqref{eq: 95} follows since all elements in $\bm{\rho}$ remain strictly positive during Bayes' update. If $\pi_1(S_0')< \lambda\pi_0(S_0')$, then
\begin{align}
  f(S_0') &= \lambda\pi_0(S_0') - \pi_1(S_0')\notag\\
    &= \lambda\big(\pi_0(S_0^*) + \rho_{i^*} \big) - \big(\pi_1(S_0^*) - \rho_{i^*} \big) \notag\\
    &= \lambda\rho_{i^*} - \big(\pi_1(S_0^*) - \lambda\pi_0(S_0^*) - \rho_{i^*} \big)\notag\\
    &< f(S_0^*), \label{eq: 96}
\end{align}
where \eqref{eq: 96} follows from the assumption that $\lambda\pi_0(S_0^*) - \pi_1(S_0^*)< -\rho_{i^*}$ and \eqref{eq: 94}. Hence, the optimality assumption of $S_0^*$ is contradicted in Case 1.

\textit{Case 2}: the partition $(S_0^*, \Omega\setminus S_0^*)$ satisfies $\lambda\pi_0(S_0^*) - \pi_1(S_0^*) > \lambda\min_{i\in S_0^*}\rho_i$. Let $i^* = \argmin_{i\in S_0^*}\rho_i$. Then,
\begin{align}
  f(S_0^*) = \lambda\pi_0(S_0^*) - \pi_1(S_0^*) > \lambda\rho_{i^*}. \label{eq: 97}
\end{align}
Consider a new subset $S_0'\triangleq S_0^*\setminus\{i^*\}$. We next show that $f(S_0') < f(S_0^*)$. There are two subcases. If $\pi_1(S_0')\ge \lambda\pi_0(S_0')$, then 
\begin{align}
  f(S_0') &= \lambda\big(\pi_1(S_0') - \lambda\pi_0(S_0') \big) \notag\\
    &= \lambda\big(\pi_1(S_0^*) + \rho_{i^*} - \lambda\pi_0(S_0^*) + \lambda\rho_{i^*} \big) \notag\\
    &= \lambda\rho_{i^*} - \lambda\big(\lambda\pi_0(S_0^*) - \pi_1(S_0^*) - \lambda\rho_{i^*} \big) \notag\\
    &< f(S_0^*), \label{eq: 98}
\end{align}
where \eqref{eq: 98} follows from the assumption that $\lambda\pi_0(S_0^*) - \pi_1(S_0^*) > \lambda\rho_{i^*}$ and \eqref{eq: 97}. If $\pi_1(S_0')< \lambda\pi_0(S_0')$, then
\begin{align}
  f(S_0') &= \lambda\pi_0(S_0') - \pi_1(S_0') \notag\\
    &= \lambda\big(\pi_0(S_0^*)-\rho_{i^*} \big) - \pi_1(S_0^*) - \rho_{i^*} \notag\\
    &< \lambda\pi_0(S_0^*) - \pi_1(S_0^*)\notag\\
    &= f(S_0^*).
\end{align}
Hence, the optimality assumption of $S_0^*$ is contradicted in Case 2.

In summary, we have shown that if the partition $(S_0^*, \Omega\setminus S_0^*)$ does not meet \eqref{eq: SED encoder for BAC}, the optimality assumption of $S_0^*$ will be contradicted. Therefore, the partition $(S_0^*, \Omega\setminus S_0^*)$ must satisfy \eqref{eq: SED encoder for BAC}. This concludes the proof of Theorem \ref{theorem: 4}.

\subsection{Proof of Theorem \ref{theorem: 5}}\label{subsec: proof of theorem 5}

Let us write $\pi_0^{(s)}$ and $\pi_1^{(s)}$ to denote the probabilities of $S_0^{(s)}$ and $S_1^{(s)}$ at iteration $s$, $s = 1, 2,\dots, M$. We prove Theorem \ref{theorem: 5} by induction.

\textit{Base case}: For $s = 1$, $\pi_0^{(1)} = \rho_{j_1}$ and $\pi_1^{(1)} = 0$. Clearly,
\begin{align}
  \lambda\pi_0^{(1)} - \pi_1^{(1)} = \lambda\rho_{j_1}\in[0, \lambda\rho_{j_1}].
\end{align}
Hence, $\pi_0^{(1)}$ and $\pi_1^{(1)}$ meet the condition in \eqref{eq: SED encoder for BAC}.

\textit{Inductive step}: Assume that for $s = k$, \eqref{eq: SED encoder for BAC} holds for $\pi_0^{(k)}$ and $\pi_1^{(k)}$. We will show that \eqref{eq: SED encoder for BAC} will also hold for $\pi_0^{(k+1)}$ and $\pi_1^{(k+1)}$. There are two cases.

\textit{Case 1}: $\pi_1^{(k)} \ge \lambda\pi_0^{(k)}$. According to Algorithm \ref{algorithm: 2}, $\pi_0^{(k+1)} = \pi_0^{(k)} + \rho_{j_{k+1}}$ and $\pi_1^{(k+1)} = \pi_1^{(k)}$. Meanwhile, $\rho_{j_{k+1}} = \min_{i\in S_0^{(k+1)}}\rho_i$ and $\min_{i\in S_1^{(k)}}\rho_i = \min_{i\in S_1^{(k+1)}}\rho_i $. Therefore,
\begin{align}
    \lambda\pi_0^{(k+1)} - \pi_1^{(k+1)} &= \big(\lambda\pi_0^{(k)} - \pi_1^{(k)}\big) + \lambda\rho_{j_{k+1}}\notag\\
      &\le \lambda\min_{i\in S_0^{(k+1)}}\rho_i,
\end{align}
and
\begin{align}
  \lambda\pi_0^{(k+1)} - \pi_1^{(k+1)} &= \big(\lambda\pi_0^{(k)} - \pi_1^{(k)}\big) + \lambda\rho_{j_{k+1}}\notag\\
    &\ge -\min_{i\in S_1^{(k+1)}}\rho_i.
\end{align}
Hence, \eqref{eq: SED encoder for BAC} holds for $\pi_0^{(k+1)}$ and $\pi_1^{(k+1)}$ in Case 1.

\textit{Case 2}: $\pi_1^{(k)} < \lambda\pi_0^{(k)}$. According to Algorithm \ref{algorithm: 2}, $\pi_0^{(k+1)} = \pi_0^{(k)}$ and $\pi_1^{(k+1)} = \pi_1^{(k)}  + \rho_{j_{k+1}}$. Meanwhile, $\rho_{j_{k+1}} = \min_{i\in S_1^{(k+1)}}\rho_i$ and $\min_{i\in S_0^{(k)}}\rho_i = \min_{i\in S_0^{(k+1)}}\rho_i $. Therefore,
\begin{align}
    \lambda\pi_0^{(k+1)} - \pi_1^{(k+1)} &= \big(\lambda\pi_0^{(k)} - \pi_1^{(k)} \big) - \rho_{j_{k+1}} \notag\\
      &\le \lambda \min_{i\in S_0^{(k+1)}}\rho_i,
\end{align}
and
\begin{align}
   \lambda\pi_0^{(k+1)} - \pi_1^{(k+1)} &= \big(\lambda\pi_0^{(k)} - \pi_1^{(k)} \big) - \rho_{j_{k+1}}\notag\\
      &> -\min_{i\in S_1^{(k+1)}}\rho_i.
\end{align}
Hence, \eqref{eq: SED encoder for BAC} holds for $\pi_0^{(k+1)}$ and $\pi_1^{(k+1)}$ in Case 2.

In summary, \eqref{eq: SED encoder for BAC} holds for $\pi_0^{(k+1)}$ and $\pi_1^{(k+1)}$ at iteration $s = k + 1$. Therefore, when the algorithm terminates, a two-way partition of $\Omega$ will be formed and the corresponding $\pi_0^{(M)}$ and $\pi_1^{(M)}$ will satisfy \eqref{eq: SED encoder for BAC}. This completes the proof of Theorem \ref{theorem: 5}.

\subsection{Proof of Lemma \ref{lemma: first supporting lemma}} \label{subsec: proof of lemma 4}

The proof of Lemma \ref{lemma: first supporting lemma} includes a construction of a surrogate submartingale and an application of the variant of Doob's optional stopping theorem. 

Let $x_i$ be the input symbol for $\theta = i$ at time $t+1$ and define $\bar{x}_i = 1 - x_i$. Following the derivation of \eqref{eq: extrinsic prob}, for $Y_{t+1} = y$,
\begin{align}
  U_i(t+1) &= \log\frac{\rho_i(t+1)}{1 - \rho_i(t+1)} \\
    &= U_i(t) + \log\frac{P_{Y|X}(y|x_i)}{\sum_{x\in\X} \tilde{\pi}_{x,i}(t)P(y|x)},
\end{align}
where $\tilde{\pi}_{x, i}(t)$, $x\in\{0, 1\}$, is the extrinsic probabilities defined in \eqref{eq: extrinsic 1} and \eqref{eq: extrinsic 2}. For brevity, let us define the instantaneous step size
\begin{align}
  w_i(t, y) \triangleq \log\frac{P_{Y|X}(y|x_i)}{\sum_{x\in\X} \tilde{\pi}_{x,i}(t)P(y|x)}.
\end{align}
From previous analysis in Section \ref{subsec: proof of lemma 3}, we showed in \eqref{eq: 39} that with the SED encoding rule, 
\begin{align}
  \E[W_i(t, Y)|\F_t] &\ge C.
\end{align}
where $C$ is the capacity of the BSC$(p)$.

Here, we seek a \emph{surrogate submartingale} $U_i'(t)$ satisfying the following two conditions:
\begin{itemize}
  \item[1).] $\forall t\ge0$ and $\forall y^t$, $U_i'(t)\le U_i(t)$ with $U_i'(0) = U_i(0)$;
  \item[2).] $\E[U_i'(t)|\F_t] = U_i'(t) + C$.
\end{itemize}
The motivation behind condition 1 is explained below. Let
\begin{align}
  \nu_i' \triangleq \min\{t:  U_i'(t)\ge 0\}.
\end{align}
Thus, Condition 1 implies that $\nu_i\le \nu_i'$.

\textit{Construction of $\{U_i'(t)\}_{t=0}^\infty$}: Let $U_i'(0) = U_i(0)$. For $t\ge 0$ and $Y_{t+1} = y$,
\begin{align}
  U_i'(t+1) \triangleq U_i'(t) + w_i'(t, y),
\end{align}
where $w_i'(t,y)$ is defined as
\begin{align}
    w_i'(t, x_i) &\triangleq \log2P_{Y|X}(x_i|x_i)\notag\\
      &\phantom{=\,}-\frac{P_{Y|X}(\bar{x}_i|x_i) }{P_{Y|X}(x_i|x_i)}\log\frac{1/2}{\sum_{x\in\X} \tilde{\pi}_{x,i}(t)P(\bar{x}_i|x) } \\
    w_i'(t, \bar{x}_i) &\triangleq \log2P_{Y|X}(\bar{x}_i|x_i)+\log\frac{1/2}{\sum_{x\in\X} \tilde{\pi}_{x,i}(t)P(\bar{x}_i|x) }.
\end{align}
Compared with $w_i(t, y)$, the only distinction lies in $w_i'(t, x_i)\ne w_i(t, x_i)$. We now show that this construction of $\{U_i'(t)\}_{t=0}^\infty$ indeed satisfies the two conditions aforementioned. First,
\begin{align}
  &\E[W_i'(t, Y)|\F_t]  \notag\\
  &= P_{Y|X}(x_i|x_i)w_i'(t, x_i) + P_{Y|X}(\bar{x}_i|x_i)w_i'(t, \bar{x}_i) \\
  &= P_{Y|X}(x_i|x_i)\log2P_{Y|X}(x_i|x_i) \notag\\
  &\phantom{=\,}+ P_{Y|X}(\bar{x}_i|x_i)\log2P_{Y|X}(\bar{x}_i|x_i)\\
  &= C.
\end{align}
This implies that $\{U_i'(t)\}_{t=0}^\infty$ is a submartingale satisfying Condition 2. Specifically, $\E[U_i'(t+1)|\F_t]=U_i'(t) + C$. 

Next, we show Condition 1 also holds for $\{U_i'(t)\}_{t=0}^\infty$. Note that the only difference between $w_i(t, y)$ and $w_i'(t, y)$ is when $y = x_i$, thus, it suffices to show $w_i'(t, x_i)\le w_i(t, x_i)$. Indeed,
\begin{align}
  &w_i(t, x_i)\notag\\
  &=\log2P_{Y|X}(x_i|x_i) + \log\frac{1/2}{\sum_{x\in\X} \tilde{\pi}_{x,i}(t)P(x_i|x)}\notag\\
  &\ge  \log2P_{Y|X}(x_i|x_i) {+} \frac{P_{Y|X}(\bar{x}_i|x_i) }{P_{Y|X}(x_i|x_i)} \log\frac{1/2}{\sum_{x\in\X} \tilde{\pi}_{x,i}(t)P(x_i|x)} \notag\\
  &\ge \log2P_{Y|X}(x_i|x_i) {+} \frac{P_{Y|X}(\bar{x}_i|x_i) }{P_{Y|X}(x_i|x_i)} \log\frac{\sum_{x\in\X} \tilde{\pi}_{x,i}(t)P(\bar{x}_i|x)}{1/2}\label{eq: ineq123}\\
  &= w_i'(t, x_i),
\end{align}
where \eqref{eq: ineq123} follows from the inequality below. Let us use the shorthand notation $\pi_{x, i} = \pi_{x,i}(t)$, $p = P_{Y|X}(\bar{x}|x)$ and $q = P_{Y|X}(x|x)$. Then,
\begin{align}
  &(\tilde{\pi}_{x_i, i}q+\tilde{\pi}_{\bar{x}_i, i}p)(\tilde{\pi}_{x_i, i}p+\tilde{\pi}_{\bar{x}_i, i}q)\notag\\
  &= -(q-p)^2\tilde{\pi}_{x_i,i}^2 + (q-p)^2\tilde{\pi}_{x_i,i} + pq \\
  &\le \frac14(q+p)^2 \\
  &= \frac14
\end{align}
with equality if and only if $\tilde{\pi}_{x_i, i} = 1/2$. Thus, $w_i'(t, y)\le w_i(t, y)$ for $y\in\{0, 1\}$ and Condition 1 follows.

Finally, we apply Lemma \ref{lemma: variant of Doob} to the surrogate submartingale $\{U_i'(t)\}_{t=0}^\infty$ to obtain an upper bound on $\E[\nu_i']$. Observe that for any $t\ge 0$ and $y^t$,
\begin{align}
  |w_i'(t, y)| &\le |w_i(t,y)|\le C_2.
\end{align}
Hence, the conditions in Lemma \ref{lemma: variant of Doob} are met.

\begin{figure*}[t]
\centering
\includegraphics[width=0.9\textwidth]{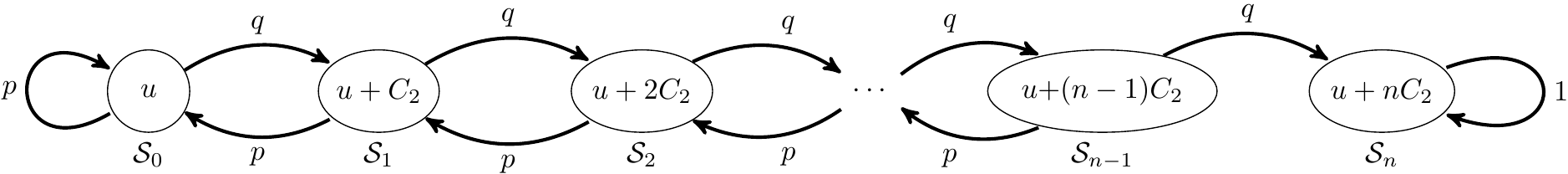}
\caption{An example of the generalized Markov chain with initial value $u$, $u\in[0, C_2)$, assuming that $U_i(t)$ arrives at $u$ when crossing threshold $0$ and remains nonnegative all the time. The value beside each branch denotes the transition probability. The value inside the $j$-th circle represents the unique active value in $\setS_j$, $1\le j\le n$.}
\label{fig: generalized Markov chain}
\end{figure*}

Consider a normalized sequence $\{\eta(t)\}_{t=0}^\infty$ defined as
\begin{align}
  \eta(t) \triangleq \frac{U_i'(t)}{C} -t.
\end{align}
It is straightforward to show that $\{\eta(t)\}_{t=0}^\infty$ is a martingale with a bounded difference $|\eta(t+1)-\eta(t)|$. Therefore, by Lemma \ref{lemma: variant of Doob},
\begin{align}
  \frac{U_i(0)}{C} &=  \eta(0)\notag\\
     &\le \E[\eta(\nu_i')] \notag\\
    &= \frac{\E[U_i'(\nu_i') - U_i'(\nu_i'-1)] + \E[U_i'(\nu_i'-1)]}{C}-\E[\nu_i'] \notag\\
    &\le \frac{w_i'(t, x_i) + 0}{C} -\E[\nu_i'] \label{eq: ineq130} \\
    &\le \frac{(1/q)\log2q}{C} -\E[\nu_i']  \label{eq: ineq131}
\end{align}
where \eqref{eq: ineq130} follows from the fact that $U_i'(t)$ has to cross the threshold $0$ from $t=\nu_i'-1$ to $t=\nu_i'$ and that $w_i'(t,x_i)$ is the only positive, instantanenous step size, \eqref{eq: ineq131} follows from the fact that
\begin{align}
  w_i'(t, x_i) &\le \log2q -\frac{p}{q}\log\frac{1/2}{q} = \frac{\log2q}{q}.
\end{align}
Combining \eqref{eq: ineq131} with the fact that $\nu_i\le \nu_i'$,
\begin{align}
  \E[\nu_i]\le \E[\nu_i'] &\le \frac{\log2q}{qC} - \frac{\log\frac{1/M}{1-1/M}}{C} \\
    &< \frac{\log M}{C} + \frac{\log2q}{qC}.
\end{align}
This completes the proof of Lemma \ref{lemma: first supporting lemma}.


\subsection{Proof of Lemma \ref{lemma: second supporting lemma}} \label{subsec: proof of lemma 5}

The proof requires several steps. First, we show that when $U_i(t)\ge 0$, the behavior of $U_i(t)$ can be modeled as a Markov chain with a fallback self loop. This self loop represents the probability that $U_i(t)$ first falls back to the communication phase and eventually returns to the confirmation phase. Next, the problem of solving $\E[\tau_i - \nu_i | \theta = i, U_i(\nu_i) = u]$ can be formulated as the expected first-passage time from the initial state to the terminating state on a generalized Markov chain.

Let $q \triangleq 1 - p$. For BSC$(p)$, $p\in(0, 1/2)$, by Fact \ref{fact: 3}, $C_2 = \log(q/p)$ and $C_1 = (q-p)C_2$. In the following analysis, we fix $\theta = i\in\Omega$ unless otherwise specified.

Recall that with the SED encoding, the one-step update for $U_i(t)$ when $U_i(t)\ge 0$ is given by \eqref{eq: instantaneous update}. In the case of BSC$(p)$, we have 
\begin{align}
  U_i(t+1) = U_i(t) + W, \label{eq: one-step transfer}
\end{align} 
where $W  = C_2$ with probability $P_{Y|X}(1|1)=q$ and $W = -C_2$ with probability $P_{Y|X}(0|1) = p$. Assume that $U_i(\nu_i) = u \in[0, C_2)$. Clearly, the behavior of $U_i(t)$ is a Markov chain with initial value $u$, provided that $U_i(t)\ge u$ for all $t\ge \nu_i$.

Unfortunately, the above Markov chain is too simple to capture the reality. First, $U_i(t)$ can fall back to the communication phase (i.e., $U_i(t) < 0$) at some $t' > \nu_i$. Second, if $U_i(t)$ falls back and then returns to the confirmation phase, the value at which $U_i(t)\ge 0$ might be different from $u$.

Nevertheless, we make two important observations. First, the prior probability that $U_i(t)$ falls back to communication phase is $p$. Given that $U_i(t)$ falls back, the conditional probabiliy that $U_i(t)$ eventually returns to the confirmation phase is $1$ (since $\tau_i$ is a.s. finite). Hence, the transition probability from the initial state value $u$ at which $U_i(t)$ falls back to another initial state value $u'$ at which $U_i(t)$ returns is $p$. Second, assume $u$ is the initial value when $U_i(t)$ first enters the confirmation phase. By \eqref{eq: one-step transfer}, the subsequent values that $U_i(t)$ assumes are of the form $u + jC_2$, $j=0,1,\dots$, provided that $U_i(t)\ge 0$ all the time. These observations motivate the definition of a \emph{generalized Markov chain}.

\begin{definition}
  Let $\setS_0 = [0, C_2)$ represent the set of values of $U_i(t)$ when transitioning from below $0$ to above $0$ for the first time. Let $n  \triangleq \lceil \log\frac{1-\epsilon}{\epsilon}/C_2 \rceil$. Define $\setS_j\triangleq [jC_2, jC_2 + C_2)$, $1\le j\le n$. The generalized Markov chain consists of a sequence of states $\setS_0, \setS_1, \dots, \setS_n$ satisfying
  \begin{align}
    \Prob\Brace{\setS_{j+1}|\setS_j} &\triangleq P_{V|U}(u+C_2| u),\ u\in\setS_j,\  0\le j\le n-1,  \notag\\
    \Prob\Brace{\setS_{j-1}|\setS_j} &\triangleq P_{V|U}(u-C_2| u),\ u\in\setS_j,\  1\le j\le n, \notag\\
    \Prob\Brace{\setS_{0}|\setS_0} &\triangleq P(V\in\setS_0| U = u),\ u\in\setS_0, \notag\\
    \Prob\Brace{\setS_{n}|\setS_n} &\triangleq 1, \notag
  \end{align} 
  where if $u\in\setS_0$, $P(V\in\setS_0| U = u) = p$ and $P(V = u+C_2| U = u) = q$. If $u\in \setS_j$, $j\ge 1$, $P(V = u + C_2|U = u) = q$ and $P(V = u + C_2|U = u) = p$.
\end{definition}

The distinction between the generalized Markov chain and a regular Markov chain discussed above is that each state is an interval rather than a single value. However, as soon as $U_i(t)\ge 0$, only a single value in each set $\setS_j$ remains active and is uniquely determined by the initial value in $\setS_0$. Specifically, if the initial value is $u$, then the only active value in $\setS_j$ is given by $u+jC_2$, $1\le j\le n$. For this reason, each state $\setS_j$, albeit defined as an interval, resembles a ``single value'', and one can directly define transition probabilities between two consecutive states. Fig. \ref{fig: generalized Markov chain} illustrates an example of the generalized Markov chain with initial value $u\in[0, C_2)$.

Let us consider a new stopping time
\begin{align}
  \tau_i^* \triangleq \min\Brace{t: \left\lfloor\frac{U_i(t)}{C_2} \right\rfloor \ge \left\lceil \frac{\log\frac{1-\epsilon}{\epsilon}}{C_2} \right\rceil }. \label{eq: new stopping time}
\end{align}
By definition, $\tau_i^*$ is independent from the initial value $U_i(t) - \lfloor U_i(t)/C_2 \rfloor C_2$ and is achieved whenever $U_i(t)$ enters $\setS_n$ for the first time. Moreover, 
\begin{align}
  \frac{U_i(\tau_i^*)}{C_2} \ge \left\lfloor\frac{U_i(\tau_i^*)}{C_2} \right\rfloor \ge \left\lceil \frac{\log\frac{1-\epsilon}{\epsilon}}{C_2} \right\rceil \ge \frac{\log\frac{1-\epsilon}{\epsilon}}{C_2}.
\end{align}
Hence, by definition of $\tau_i$ in \eqref{eq: def tau_i}, we obtain
\begin{align}
    \tau_i \le \tau_i^*. 
\end{align}
This implies that
\begin{align}
  \E[\tau_i - \nu_i | \theta = i, U_i(\nu_i) = u]\le \E[\tau_i^* - \nu_i | \theta = i, U_i(\nu_i) = u]. \label{eq: ineq139}
\end{align}
Note that $\E[\tau_i^* - \nu_i | \theta = i, U_i(\nu_i) = u]$ represents the expected first-passage time from initial state $u\in\setS_0$ when $U_i(t)$ first crosses threshold $0$ to state $\setS_n$. In Appendix \ref{appendix: time of first passage analysis}, the time of first passage analysis reveals that
\begin{align}
  &\E[\tau_i^* - \nu_i | \theta = i, U_i(\nu_i) = u] \notag\\
  &= \frac{n}{1-2p}+\frac{p}{1-2p}\Paren{1 - \Big(\frac{p}{q}\Big)^n }\left(\Delta_0 - \frac{2q}{1-2p} \right) \\
  &= \frac{nC_2}{C_1} + \frac{2^{-C_2}}{1-2^{-C_2}}(1 - 2^{-nC_2})\left(\Delta_0 -1 -\frac{C_2}{C_1} \right), \label{eq: time_first_passage}
\end{align}
where $\Delta_0$ represents the expected self loop time of $U_i(t)$ from $\setS_0$ to $\setS_0$. Assume that after fallback, $U_i(t) = u - C_2 < 0$. Following the proof of Lemma \ref{lemma: first supporting lemma} in Section \ref{subsec: proof of lemma 4}, we immediately obtain,
\begin{align}
  \Delta_0 &\le 1 + \frac{(1/q)\log2q}{C} - \frac{u-C_2}{C}\\
    &= 1 + \frac{(1/q)\log2q + C_2 - u}{C}\\
    &\le  1+ \frac{(1/q)\log2q + C_2 }{C}. \label{eq: upper bound Delta_0}
\end{align}
Substituting \eqref{eq: upper bound Delta_0} into \eqref{eq: time_first_passage} yields
\begin{align}
  &\E[\tau_i^* - \nu_i | \theta = i, U_i(\nu_i) = u] \notag\\
  &\le \frac{nC_2}{C_1} + C_2 2^{-C_2}\frac{1 - 2^{-nC_2}}{1-2^{-C_2}}\left(\frac{1+\frac{\log2q}{qC_2}}{C}-\frac{1}{C_1} \right).
\end{align}

Using $n = \lceil \log\frac{1-\epsilon}{\epsilon}/C_2 \rceil \le \log\frac{1-\epsilon}{\epsilon}/C_2 + 1$, we obtain the desired upper bound
\begin{align}
  &\E[\tau_i^* - \nu_i | \theta = i, U_i(\nu_i) = u] \notag\\
  &\le \frac{\log\frac{1-\epsilon}{\epsilon} {+} C_2 }{C_1} + C_22^{-C_2}\frac{1 - \frac{\epsilon}{1-\epsilon}2^{-C_2}}{1 - 2^{-C_2}}\left(\frac{1{+}\frac{\log2q}{qC_2}}{C} - \frac{1}{C_1} \right). 
\end{align}
Invoking \eqref{eq: ineq139} completes the proof of Lemma \ref{lemma: second supporting lemma}.


\section{Numerical Simulation} \label{sec: numerical simulation}

In this section, we simulate the proposed SED encoder for a regularized BAC$(p_0, p_1)$ with feedback. In the case of BSC, we will compare the empirical rate with the achievability bound given by our results and previous work. By \eqref{eq: achievable rate}, we can compute the empirical rate achieved by the SED encoder. Using the non-asymptotic upper bound on $\E[\tau]$, we can also obtain the associated achievability bound (i.e., lower bound) on rate.

\subsection{Simulation Results on BAC with Feedback}

Let the target error probability $\epsilon = 10^{-3}$. Consider the BAC$(0.03, 0.22)$ with feedback. Using Facts \ref{fact: 1} and \ref{fact: 3}, one can compute
\begin{align}
  C = 0.5, \quad C_1 = 3.1954, \quad C_2 = 4.7.
\end{align}

\begin{figure}[t]
\centering
\includegraphics[width=0.45\textwidth]{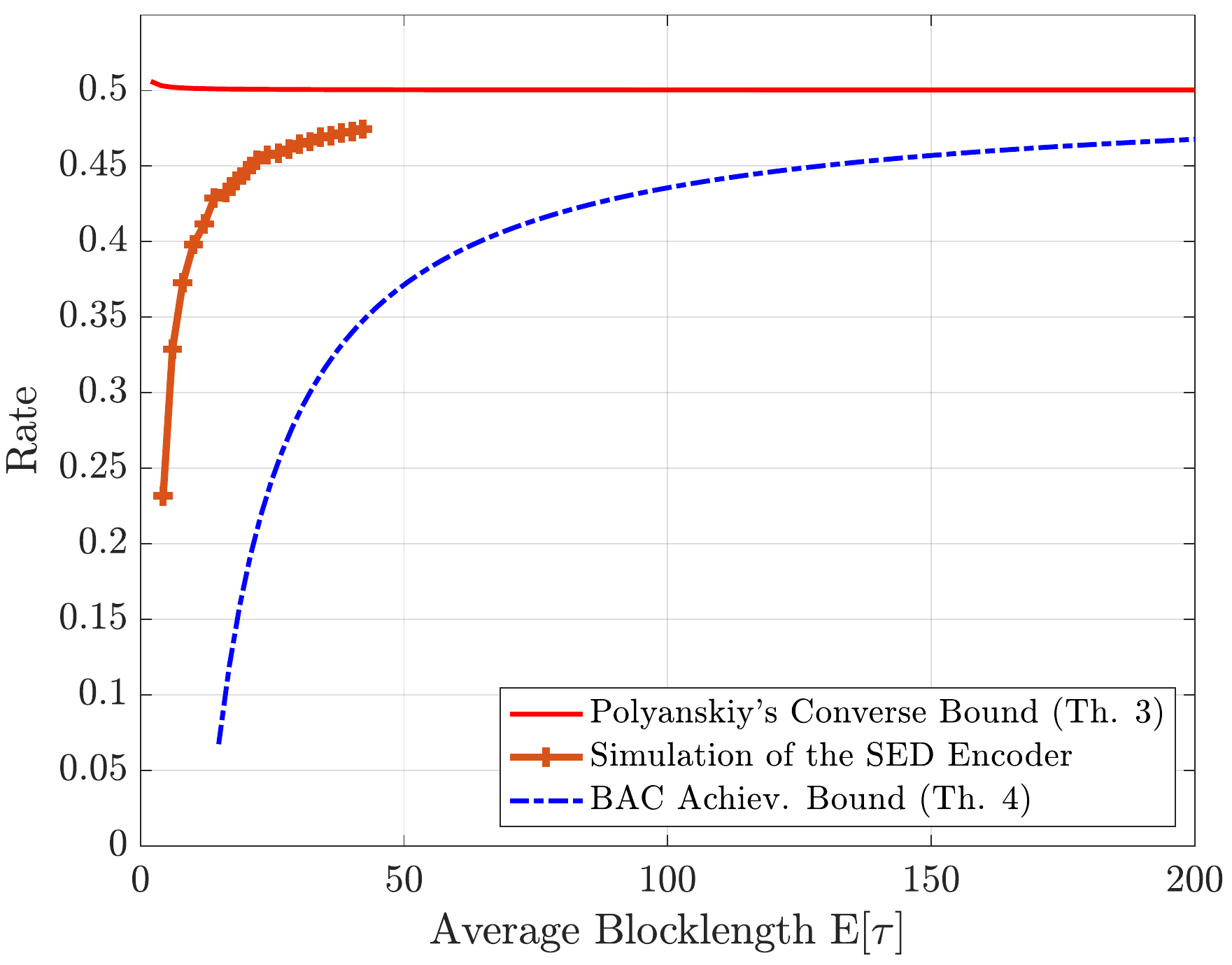}
\caption{The rate as a function of average blocklength over the BAC$(0.03, 0.22)$ with noiseless feedback. Target error probability $\epsilon = 10^{-3}$.}
\label{fig: sim on BAC}
\end{figure}

Fig. \ref{fig: sim on BAC} shows the simulated rate vs. average blocklength of the SED encoder over the BAC$(0.03, 0.22)$ with feedback, along with the achievability bound derived from Theorem \ref{theorem: BAC bound}. Since the SED encoder has an exponential complexity in message length, we were only able to simulate the message length from $k = 1$ to $k = 20$ bits. We see in Fig. \ref{fig: sim on BAC} that the actual performance of the SED encoder is much better than our achievability bound, implying that there is still room for improvement.

\subsection{Simulation Results on BSC with Feedback}

Consider the target error probability $\epsilon = 10^{-3}$ and the BSC$(0.11)$ with feedback. Using Facts \ref{fact: 1} and \ref{fact: 3},
\begin{align}
  C = 0.5,\quad C_1 = 2.3527, \quad C_2 = 3.0163.
\end{align}
One can verify that this setting satisfies the technical conditions in \cite{Naghshvar2015}. Thus, by Theorem \ref{theorem: 1} of Naghshvar \emph{et al.},
\begin{align}
  \E[\tau]\le \frac{\log M + \log\log M + 3.317}{0.5} + 4.6609 + 5341.38,
\end{align}
which turns out to be a much loose bound. The corresponding achievability bound even falls out of the average blocklength region of interest, thus is omitted from the simulation plot.

\begin{figure}[t]
\centering
\includegraphics[width=0.47\textwidth]{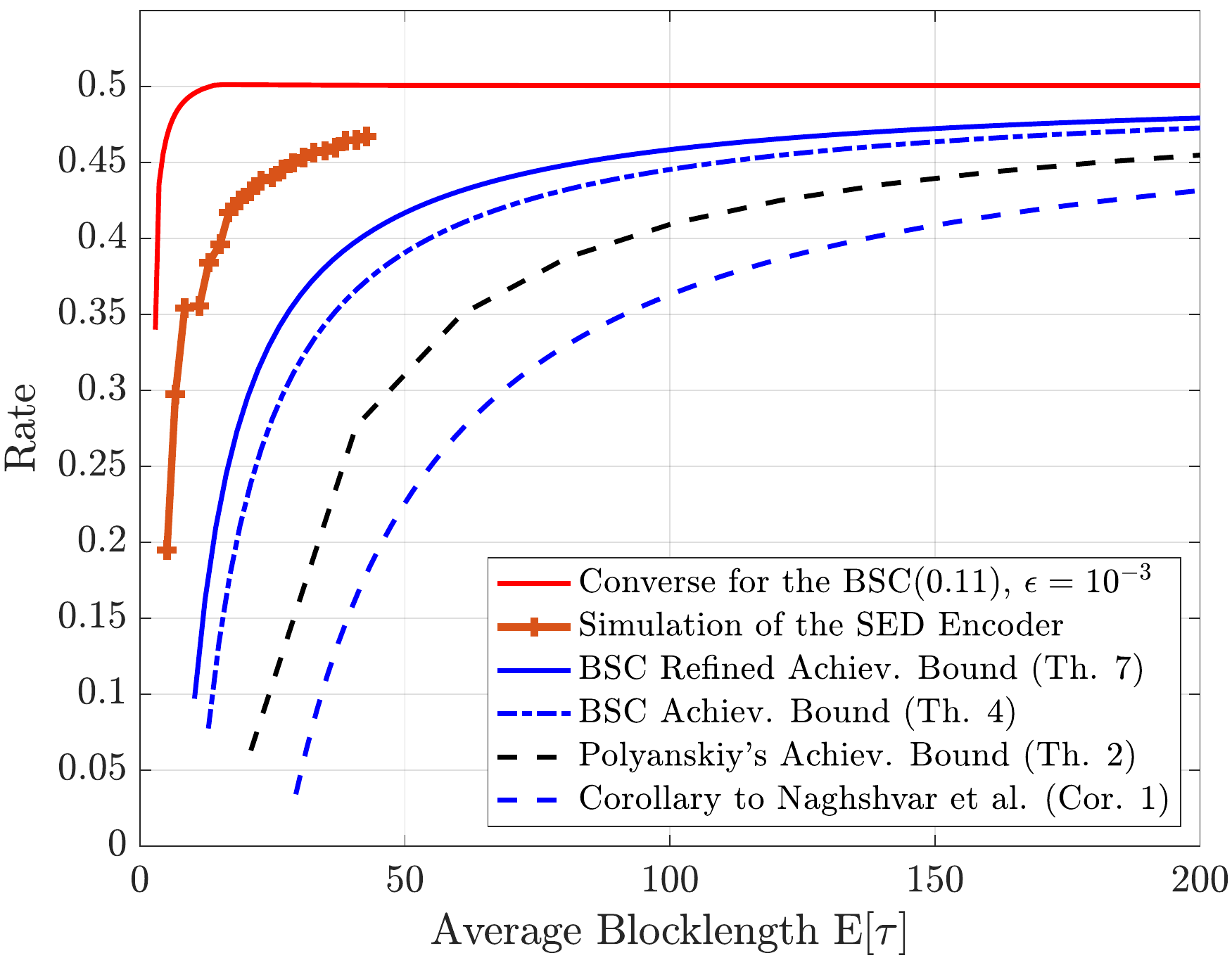}
\caption{The rate as a function of average blocklength over the BSC$(0.11)$ with noiseless feedback. Target error probability $\epsilon = 10^{-3}$.}
\label{fig: sim on BSC}
\end{figure}

Fig. \ref{fig: sim on BSC} shows the numerical comparison of the upper and lower bounds, and the actual performance of the SED encoder for the BSC with crossover probability $p = 0.11$ and $\epsilon = 10^{-3}$. Due to the exponential encoding complexity, we were only able to simulate message lengths from $k = 1$ to $k = 20$. Despite that Corollary \ref{corollary: 1} is a better result compared to Theorem \ref{theorem: 1}, the resulting achievability bound still falls beneath Polyanskiy's achievability bound on rate for a system limited to stop feedback. In contrast, our BSC achievability bound from Theorem \ref{theorem: BAC bound} exceeds Polyanskiy's achievability result in \cite{Polyanskiy2011}. Indeed, this should be expected since a system that employs full, noiseless feedback should perform better than a system with stop feedback. In particular, the refined achievability bound from Theorem \ref{theorem: BSC bound} is a further improvement compared to Theorem \ref{theorem: BAC bound}.

\section{Implications on the Reliability Function}\label{sec: reliability function}

In this section, we show that our SED encoder for the regularized BAC with feedback attains both capacity and Burnashev's optimal error exponent.

Let $\mathfrak{c}$ be a variable-length coding scheme such that for each positive number $l$, one out of $M_{\mathfrak{c}_l}$ equiprobable messages is transmitted at an error probability $P_{e,\mathfrak{c}}$ and with an average blocklength $\E_{\mathfrak{c}_l}[\tau]$. We say that the scheme $\mathfrak{c}$ achieves rate $R$ if for any small numbers $\delta > 0$, $\epsilon\in[0, 1)$ and all sufficiently large $l$, the following three conditions hold:
\begin{subequations}
\begin{align}
  P_{e,\mathfrak{c}_l} &\le \epsilon, \\
  M_{\mathfrak{c}_l} &\ge 2^{l(R-\delta)}, \label{eq: 154b} \\
  \E_{\mathfrak{c}_l}[\tau] &\le l. \label{eq: 154c}
\end{align}
\end{subequations}
Furthermore, if the scheme $\mathfrak{c}$ satisfies \eqref{eq: 154b}, \eqref{eq: 154c} and a stronger condition
\begin{align}
  P_{e,\mathfrak{c}_l} &\le 2^{-l(E - \delta)},
\end{align}
for some positive real number $E$, then we say the scheme $\mathfrak{c}$ achieves error exponent $E$ at rate $R$.

We invoke a general result from \cite{Naghshvar2015} to show our claim.
\begin{lemma}[Lemma 4, \cite{Naghshvar2015}]\label{lemma: 11}
  Suppose that we have a VLF coding scheme $\mathfrak{c}$ that for each message size $M > 0$ and each positive $\epsilon > 0$, satisfies $P_{e,\mathfrak{c}}\le\epsilon$ with expected stopping time
  \begin{align}
  \E_{\mathfrak{c}}[\tau]\le\Paren{\frac{\log M}{R_{\min}} + \frac{\log\frac{1}{\epsilon}}{E_{\min} } }\big(1 + o(1)\big)
  \end{align}
  for some positive numbers $E_{\min}$ and $R_{\min}$, where $o(1)\to 0$ as $\epsilon\to0$ or $M\to\infty$. Then, the scheme $\mathfrak{c}$ can achieve any rate $R\in[0, R_{\min}]$ with error exponent $E$, if
  \begin{align}
    E \le E_{\min}\Paren{1 - \frac{R}{R_{\min} } }.
  \end{align}
\end{lemma}
Observe that in Theorem \ref{theorem: BAC bound} and Theorem \ref{theorem: BSC bound}, both upper bounds can be relaxed and written in the form of 
\begin{align}
\frac{\log M}{C} + \frac{\log\frac{1}{\epsilon}}{C_1} + \frac{K(C, C_1, C_2)}{CC_1}, 
\end{align}
where $K(C, C_1, C_2)$ is a constant that only relies on $C, C_1, C_2$. Hence, for sufficiently large $M$ or sufficiently small $\epsilon$,
\begin{align}
  \frac{K(C, C_1, C_2)}{CC_1} \le \frac{C_1\log M + C\log\frac{1}{\epsilon}}{CC_1} = \frac{\log M }{C} + \frac{\log\frac{1}{\epsilon}}{C_1}. \notag
\end{align}
implying that the non-asymptotic upper bounds in both Theorem \ref{theorem: BAC bound} and Theorem \ref{theorem: BSC bound} meet the condition in Lemma \ref{lemma: 11}. Therefore, the SED encoding scheme can achieve any rate $R\in[0, C]$ with error exponent
\begin{align}
  E\le C_1\Paren{1- \frac{R}{C}},
\end{align}
thus the claim is proved.

\section{Conclusion} \label{sec: conclusion}

In this paper, we proposed a generalized SED encoder for the class of binary asymmetric channels with full, noiseless feedback. For a BSC with feedback, this generalized SED encoder a relaxation of Naghshvar \emph{et al.}'s SED encoder.   This paper develops a non-asymptotic upper bound on the average blocklength of the VLF code associated with the generalized SED encoding rule. For the example of the BSC with capacity $1/2$, the corresponding lower bound on achievable rate for a system with full feedback is above  Polyanskiy's achievability bound for a system limited to stop-feedback codes.  In summary, the SED encoding rule is a powerful tool that helps facilitate new achievability bounds. 

The theoretical development utilized the concept of extrinsic probabilities introduced by Naghshvar \emph{et al.} to connect the generalized SED encoder to the corresponding non-asymptotic upper bound on average blocklength. The partial ordering $\pi_1^*\le \pi_0^*$ for a regularized BAC implies that transmitting symbol $1$ achieves the maximum relative entropy $C_1$.  These observations facilitate the generalized SED encoder.  However, it remains open as to whether these observations also hold for a general binary-input channel with feedback.

\appendices
\section{Proof of Fact \ref{fact: 1}} \label{appendix: proof of fact 1}

Let $(\pi_0, 1-\pi_0)$ be an input distribution to a BAC$(p_0, p_1)$. Hence, $Y$ is also a binary random variable with
\begin{align}
  \Prob\Brace{Y = 0} = \pi_0(1 - p_0) + (1 - \pi_0)p_1.
\end{align}
Therefore, the mutual information $I(\pi_0)$ between $X$ and $Y$ is given by
\begin{align}
  &I(\pi_0) \notag\\
  &= h\big(\pi_0(1 - p_0) + (1 - \pi_0)p_1 \big) - \pi_0h(p_0) - (1 - \pi_0)h(p_1) \notag\\
  &= h\big(\pi_0(1-p_0-p_1)+p_1 \big)-\pi_0\big(h(p_0)-h(p_1) \big) - h(p_1). \label{eq: MI}
\end{align}
Since mutual information $I(\pi_0)$ is strictly concave in $\pi_0\in(0, 1)$ \cite{Cover:2006}, the optimal $\pi_0^*$ satisfies $I'(\pi_0^*) = 0$. The first derivative of $I(\pi_0)$ is given by
\begin{align}
  I'(\pi_0) &= (1-p_0-p_1)\log\Paren{\frac{1}{\pi_0(1-p_0-p_1)+p_1}-1 } \notag\\
    &- \big(h(p_0)-h(p_1) \big).
\end{align}
Clearly, $I'(\pi_0)$ is a monotonically decreasing function in $\pi_0\in(0, 1)$. Let $z \triangleq 2^{\frac{h(p_0)-h(p_1)}{1-p_0-p_1}}$. By setting $I'(\pi_0) = 0$, we obtain $\pi_0^*$ in \eqref{eq: 2}. Using \eqref{eq: MI} and the relation $\pi_1^* = 1 - \pi_0^*$, we obtain capacity $C$ in \eqref{eq: 1} and $\pi_1^*$ in \eqref{eq: 3}.

To show that $p_0\in(0, 1/2)$ and $p_0\le p_1\le 1-p_0$ imply $\pi_0^*\ge 1/2$, it suffices to show that
\begin{align}
  I'\Paren{\frac12} \ge 0.
\end{align}
Note that
\begin{align}
  I'\Paren{\frac12} =& -(1-p_0-p_1)\log\Paren{\frac{1}{\frac{(1-p_1)+p_0}{2} }-1 } -h(p_0) \notag\\
    &+ h(1-p_1).
\end{align}
Therefore, it is equivalent to show that
\begin{align}
  h(1-p_1) \ge h(p_0) + (1-p_1-p_0)\log\Paren{\frac{1}{\frac{(1-p_1)+p_0}{2} }-1 }. \label{eq: 110}
\end{align}
Let us fix $p_0\in(0, 1/2)$ and define $x\triangleq 1-p_1\in[p_0, 1-p_0]$. Then, \eqref{eq: 110} simplifies to
\begin{align}
  h(x) \ge h(p_0) + (x - p_0)\log\Paren{\frac{1}{\frac{x+p_0}{2} } - 1 }. \label{eq: 111}
\end{align}

In order to show \eqref{eq: 111}, we introduce the following useful lemma.
\begin{lemma}\label{lemma: 9}
  Let $f:(0, 1)\to\R$ be convex in $(0, 1/2]$ and be concave in $[1/2, 1)$. Additionally, $f(x) = - f(1-x)$. Then, $\forall x,y\in(0, 1)$ with $x + y < 1$,
  \begin{align}
    f(x) + f(y) \ge 2f\Paren{\frac{x+y}{2}}. \label{eq: 112}
  \end{align}
\end{lemma}
\begin{IEEEproof}
  Without loss of generality, assume that $x < y$. If $y \le 1/2$, \eqref{eq: 112} directly follows from convexity of $f(x)$ in $x\in(0, 1/2]$. Now consider $y > 1/2$. Therefore,
  \begin{align}
    &f\Paren{\frac{x+y}{2}} - f(y) \notag\\
  &= f\Paren{\frac{y+x}{2}} - f\Paren{1/2} + f\Paren{1/2} - f(y) \notag \\
  &\le f\Paren{\frac{1-y+x}{2}} - f(1-y) + f\Paren{1/2} - f(y)\label{eq: 113}\\
  &= f(1/2) - f\Paren{\frac{1+y-x}{2}} \label{eq: 114}\\
  &\le f\Paren{1 - \frac{x+y}{2}} - f(1 - x) \label{eq: 115}\\
  &= f(x) - f\Paren{\frac{x+y}{2}}. \label{eq: 116}
  \end{align}
  In the above,
  \begin{itemize}
    \item \eqref{eq: 113} follows from the convexity property that for a fixed $\delta > 0$,
    \begin{align}
      f(x) - f(x+\delta) \le f(y) - f(y+\delta),\ \text{whenever } x \ge y,\notag
    \end{align}
    \item \eqref{eq: 114} and \eqref{eq: 116} follow from $f(x) = -f(1-x)$,
    \item \eqref{eq: 115} follows from the concavity property that for a fixed $\delta > 0$,
    \begin{align}
      f(x) - f(x+\delta) \le f(y) - f(y+\delta),\ \text{whenever } x \le y. \notag
    \end{align}
  \end{itemize}
  This completes the proof of Lemma \ref{lemma: 9}.
\end{IEEEproof}

We are now in a position to prove \eqref{eq: 111}. Let $g(x)\triangleq \log(1/x-1)$, $x\in[p_0, 1-p_0]$. Observe that $g(x)$ meets the conditions in Lemma \ref{lemma: 9} and $h'(x) = g(x)$. Hence, appealing to Lemma \ref{lemma: 9}, we obtain
\begin{align}
  h(x) &= h(p_0) + \int_{p_0}^x g(z) \diff z\\
    &= h(p_0) + \int_{p_0}^{\frac{x+p_0}{2}}\Big(g(z) + g(x+p_0-z) \Big)\diff z\\
    &\ge  h(p_0) + \int_{p_0}^{\frac{x+p_0}{2}}2g\Paren{\frac{x+p_0}{2}}\diff z\\
    &= h(p_0) + (x-p_0)\log\Paren{\frac{1}{\frac{x+p_0}{2}}-1}.
\end{align}
This implies that \eqref{eq: 111} indeed holds. Hence, $\pi_0^*\ge 1/2$, concluding the proof of Fact \ref{fact: 1}.


\section{Proof of Fact \ref{fact: 3}} \label{appendix: proof of fact 3}

For brevity, let us define two distributions
\begin{align}
  P &\triangleq P(Y|X=0) = [1-p_0, p_0],\\
  Q &\triangleq P(Y|X=1) = [p_1, 1-p_1].
\end{align}
Hence, it is equivalent to show that
\begin{align}
  D\big(Q \| P \big) \ge D\big(P \| Q \big). \label{eq: 123}
\end{align}
Let us define the function
\begin{align}
  &f(p_0, p_1) \triangleq D\big(Q \| P \big) - D\big(P \| Q \big)\notag\\
    &= (1-p_0+p_1)\log\frac{p_1}{1-p_0} + (1+ p_0-p_1)\log\frac{1-p_1}{p_0}.
\end{align}
The first and second derivatives with respect to $p_1$ are, respectively, given by
\begin{align}
  \frac{\partial f}{\partial p_1} &= \log\frac{p_1}{1-p_0} - \log\frac{1-p_1}{p_0} + (\log e)\Big(\frac{1-p_0}{p_1} - \frac{p_0}{1-p_1} \Big)\\
  \frac{\partial^2 f}{\partial p_1^2} &= \frac{-(\log e)(2p_1-1)(p_1-1+p_0)}{p_1^2(1-p_1)^2}\\
    &\phantom{=\,}\begin{cases}
      < 0, & \text{if } p_1\in [p_0, 1/2)\\
      \ge0, & \text{if } p_1\in [1/2, 1-p_0].
    \end{cases}
\end{align}
Hence, for a given $p_0\in(0, 1/2)$, $f(p_0, p_1)$ is concave in $p_1\in[p_0, 1/2]$ and is convex in $p_1\in[1/2, 1-p_0]$. Next, we borrow a classical result in analysis \cite{Rudin1976}.
\begin{lemma} \label{lemma: 12}
  Consider a function $\phi: I\to\R$ defined on an interval $I\triangleq [a, b]$ with $a < b$. If the first derivative $\phi'(x)$ is continuous on $I$ and the second derivative $\phi''(x)$ exists for every $x\in I^o \triangleq (a, b)$, then the following two properties hold
  \begin{itemize}
    \item[1)] if $\phi''(x)\ge 0$, $x\in I^o$, and $\phi'(x^*) = 0$ for some $x^*\in I$, then $\phi(x)\ge \phi(x^*)$ for all $x\in I$.
    \item[2)] If $\phi''(x)\le 0$, $x\in I^o$, then $\phi(x)\ge \min\{\phi(a), \phi(b)\}$ for all $x\in I$.
  \end{itemize}
\end{lemma}
By Lemma \ref{lemma: 12}, for $p_1\in[p_0, 1/2]$, due to concavity,
\begin{align}
  f(p_0, p_1)&\ge \min\{f(p_0, p_0), f(p_0, 1/2) \}\\
    & =\min\{0, f(p_0, 1/2) \}. \label{eq: 129}
\end{align}
Similarly, for $p_1\in[1/2, 1-p_0]$, due to convexity and the fact that $\frac{\partial f}{\partial p_1}\big|_{p_1=1-p_0} = 0$,
\begin{align}
  f(p_0, p_1) \ge f(p_0, 1-p_0) = 0, \label{eq: 130}
\end{align}
implying that $f(p_0, 1/2)\ge 0$. Combining this with \eqref{eq: 129} and \eqref{eq: 130}, we conclude that $f(p_0, p_1)\ge 0$ for all $p_1\in[p_0, 1-p_0]$, thus establishing \eqref{eq: 123}. This completes the proof of \eqref{eq: 7}.

Next, we prove \eqref{eq: 8}. This is equivalent to showing that $p_1(1-p_1)\ge p_0(1-p_0)$, which clearly holds when $p_0\le p_1\le 1-p_0$.

\begin{figure}[t]
\centering
\includegraphics[width=0.45\textwidth]{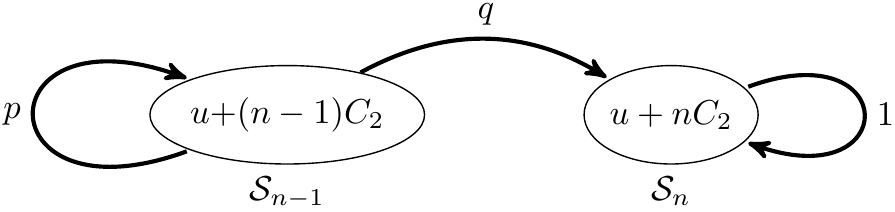}
\caption{An equivalent Markov chain from $\setS_{n-1}$ to $\setS_n$.}
\label{fig: last stage}
\end{figure}

\section{Time of First Passage Analysis} \label{appendix: time of first passage analysis}

In this section, we compute the expected first-passage time from $\setS_0$ to $\setS_n$ for the generalized Markov chain, as depicted in Fig. \ref{fig: generalized Markov chain}. Consider the general case where the self loop for state $\setS_0$ has weight $\Delta_0$ (i.e., the expected self loop time from $\setS_0$ to $\setS_0$), and all other transitions in graph has weight $1$. Let $v_i$ denote the expected first-passage time from $\setS_i$ to $\setS_n$, $0\le i\le n-1$. Our goal is to compute $v_0$, which is equal to $\E[\tau_i^* - \nu_i |\theta = i, U_i(\nu_i) = u]$.

This appendix computes $v_0$ by first simplifying the expected first-passage time node equations into an expression involving only $v_0$ and $v_{n-1}$. Characterizing the entire process to the left of $\setS_{n-1}$ as a self loop with weight $\Delta_{n-1}$ yields an explicit expression for $v_{n-1}$. This eventually produces an expression for $v_0$ that naturally decomposes into the expected first-passage time for a classical random walk plus a differential term.

\subsection{Simplifying Node Equations}

Following \cite[Chapter 4.5.1]{Gallagerbook_2013}, the node equations for the generalized Markov chain in Fig. \ref{fig: generalized Markov chain} are as follows:
\begin{subequations}\label{eq: node equations}
\begin{align}
  v_{n-1} &= 1 + pv_{n-2}, \\
  v_{n-2} &= 1 + pv_{n-3} + qv_{n-1}, \\
  v_{n-3} &= 1 + pv_{n-4} + qv_{n-2}, \\
  &\vdots \notag\\
  v_{2} &= 1 + pv_1 + qv_3, \\
  v_{1} &= 1 + pv_0 + qv_2, \\
  v_{0} &= q + pv_0 + qv_1 + p\Delta_0.
\end{align}
\end{subequations}
Summing all equations in \eqref{eq: node equations} yields
\begin{align}
  \sum_{i=0}^{n-1}v_i = n - 1 + q + \sum_{i=1}^{n-2}v_i + qv_{n-1} + 2pv_0 + p\Delta_0.
\end{align}
Solving for $v_0$ yields
\begin{align}
  v_0 &= \frac{n}{1 - 2p} + \frac{p}{1 - 2p}(\Delta_0 - v_{n-1} - 1). \label{eq: 175}
\end{align}
Thus, what remains to determine $v_0$ is to determine $v_{n-1}$.

\subsection{Expressing $v_{n-1}$ in Terms of $\Delta_0$}

\begin{figure}[t]
\centering
\includegraphics[width=0.4\textwidth]{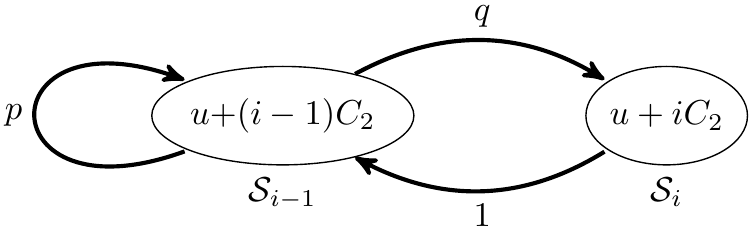}
\caption{Recursive relation between $\Delta_{i}$ and $\Delta_{i-1}$.}
\label{fig: recursive relation}
\end{figure}

In this subsection, we aim to express $v_{n-1}$ in terms of $\Delta_0$. By characterizing the entire process to the left of $\setS_{n-1}$ as a self loop with weight $\Delta_{n-1}$ and transition probability $p$, the generalized Markov chain in Fig. \ref{fig: generalized Markov chain} reduces to a two-state Markov chain as shown in Fig. \ref{fig: last stage}.
The node equation at $\setS_{n-1}$ in Fig. \ref{fig: last stage} is given by
\begin{align}
  v_{n-1} = p\Delta_{n-1} + q + pv_{n-1}.
\end{align}
Solving for $v_{n-1}$ yields
\begin{align}
  v_{n-1} = \frac{p}{q}\Delta_{n-1} + 1. \label{eq: 177}
\end{align}
Next, we develop an recursive equation to solve $\Delta_{n-1}$. Let $\Delta_i$ denote the expected self loop weight for $\setS_{i}$, $0\le i\le n-1$. Fig. \ref{fig: recursive relation} shows the transition between $\setS_{i-1}$ and $\setS_i$ conditioned on circling over $\setS_i$ once. Thus, from Fig. \ref{fig: recursive relation}, we obtain
\begin{align}
  \Delta_{i} &= 1 + \sum_{k=0}^\infty p^kq(k\Delta_{i-1} + 1) \\
    &= 2  + \frac{p}{q}\Delta_{i-1}. \label{eq: recursive formula}
\end{align}

Since \eqref{eq: recursive formula} holds for an arbitrary $i$, $0\le i\le n-1$, applying \eqref{eq: recursive formula} in a recursive manner yields
\begin{align}
  \Delta_{n-1} &= \Paren{\frac{p}{q}}^{n-1}\Delta_0 + \frac{2q}{1-2p}\Br{1 - \Paren{\frac{p}{q} }^{n-1} }. \label{eq: 180}
\end{align}
Substituting \eqref{eq: 180} into \eqref{eq: 177}, we obtain
\begin{align}
  v_{n-1} = \Paren{\frac{p}{q}}^n\Delta_0 + \frac{2p}{1-2p}\Br{1 - \Paren{\frac{p}{q} }^{n-1} } + 1. \label{eq: 181}
\end{align}

\subsection{Finding the General Expression for $v_0$}

Substituting \eqref{eq: 181} into \eqref{eq: 175},
\begin{align}
  v_0 &= \frac{n}{1-2p} + \frac{p}{1-2p}\Bigg\{ \Br{1 - \Bigg(\frac{p}{q}\Bigg)^n}\Delta_0 \notag\\
    &\phantom{=\,}- \frac{2p}{1-2p}\Br{1 - \Paren{\frac{p}{q} }^{n-1} } - 2 \Bigg\} \\
    &= \frac{n}{1-2p} + \frac{p}{1-2p}\Paren{1 - \Bigg(\frac{p}{q}\Bigg)^n }\Paren{\Delta_0 - \frac{2q}{1-2p} }. \label{eq: expression for v0}
\end{align}
This completes the derivation of $v_0$.
\begin{remark}
  For an independent and identically distributed (i.i.d.) random walk that moves forward by $1$ with probability $q$ and moves backward by $1$ with probability $p$, all $\Delta_i$'s are identical. Using \eqref{eq: recursive formula}, we obtain
  \begin{align}
    \Delta_i = \frac{2q}{1 - 2p}, \quad \forall i\in\mathbb{Z}.
  \end{align}
  Thus, \eqref{eq: expression for v0} can be thought of as the expected first-passage time for an i.i.d. random walk plus a differential term that depends on the difference between the self loop weight $\Delta_0$ of the actual random process and the self loop weight of a standard i.i.d. random walk.
\end{remark}

\section*{Acknowledgment}

The authors would like to thank Yury Polyanskiy for providing the numerical computations of the achievability and converse bounds for the BSC with feedback.

\ifCLASSOPTIONcaptionsoff
  \newpage
\fi

\bibliographystyle{IEEEtran}
\bibliography{IEEEabrv,references}

\end{document}